\documentclass[aps,prb,reprint,superscriptaddress,nofootinbib,floatfix,longbibliography]{revtex4-2}
\usepackage{amsmath,amssymb,bm,graphicx,booktabs,array}
\usepackage{xcolor}
\usepackage{placeins}
\usepackage[colorlinks=true,citecolor=blue,linkcolor=blue,urlcolor=blue]{hyperref}

\begin{document}
\title{Floquet–spin-orbit compensation and flat- and quadratic-band contact in the $\alpha$-$T_3$ lattice}
\author{Imtiaz Khan}
\affiliation{Department of Physics, Zhejiang Normal University, Jinhua 321004, China}
\affiliation{Research Center of Astrophysics and Cosmology, Khazar University, Baku, AZ1096, Azerbaijan}
\author{Muzamil Shah}
\email{muzamil@qau.edu.pk}
\affiliation{Department of Physics, Quaid-i-Azam University, Islamabad 45320, Pakistan}
\affiliation{Research Center of Astrophysics and Cosmology, Khazar University, Baku, AZ1096, Azerbaijan}
\author{Reza Asgari}
\affiliation{Department of Physics, Zhejiang Normal University, Jinhua 321004, China}
\affiliation{School of Quantum Physics and Matter, Institute for Research in Fundamental Sciences (IPM), Tehran 19395-5531, Iran}
\author{Gao Xianlong}
\email{gaoxl@zjnu.edu.cn}
\affiliation{Department of Physics, Zhejiang Normal University, Jinhua 321004, China}


\begin{abstract}
We investigate Floquet–spin-orbit compensation in the three-band $\alpha$-$T_3$ system, which interpolates between graphene at $\alpha=0$ and the dice lattice at $\alpha=1$. By tuning the interplay between an off-resonant circularly polarized optical field and intrinsic spin–orbit coupling, we identify three distinct pairwise band-degeneracy conditions in the quasienergy spectrum: (i) valley–spin degeneracy, (ii) spin–band degeneracy, and (iii) flat–dispersive band degeneracy. The flat band is not an artifact of a two-band reduction but rather an exact property of the continuum Hamiltonian at this compensation point, where the complete three-band Hamiltonian supports a momentum-independent dark state. A winding-two low-energy Hamiltonian and a finite-momentum Berry-curvature maximum, whose radius scales as the square root of the detuning, emerge from detuning this contact. We further show that the intrinsic transverse thermoelectric response encodes these spectrum patterns. The flat- and quadratic-band degeneracy can be predicted for $0<\alpha<1/2$ because the two response scales linked to the linear contacts have a monotonic ratio independent of the common spin--orbit energy scale. This allows an inverse determination of $\alpha$ and of the common spectral scale. The analytical results are confirmed by full three-band Kubo calculations, which also define the regime in which this inverse characterization is still observable. These results connect an experimentally accessible, Berry-curvature-sensitive thermoelectric response to tunable Floquet--spin--orbit band geometry.

\end{abstract}
\maketitle

\section{Introduction}\label{sec:intro}
Controlling mass terms---their sublattice pattern, their relative strength, and the band degeneracies that appear when they compensate---is a central tool of band-structure engineering in two-dimensional Dirac materials. The $\alpha$-$T_3$ lattice provides an exceptionally clean platform for this program, since a single parameter interpolates between the graphene and dice limits while keeping an exactly flat band in the spectrum. The continuum model of the $\alpha$-$T_3$ lattice is a minimal three-band extension of the honeycomb Dirac problem in which a hub site couples to two rim sublattices with unequal amplitudes. The interpolation parameter $\alpha$ changes the internal composition of the Bloch spinor without removing the two valley-centered conical bands and the intervening flat band. The graphene limit $\alpha=0$ reduces the active low-energy sector to pseudospin $1/2$, whereas $\alpha=1$ gives the dice lattice and its pseudospin-1 Dirac-Weyl structure~\cite{Bercioux2009,Raoux2014, LeeFuAng2024, PhysRevB.105.165402, Iurov2019, b4tj-5rky}. Between these endpoints the Berry phase evolves continuously with the relative rim-site weight, and this evolution appears in Landau-level quantization, magneto-optical response, Hall quantization, Klein tunneling, and ballistic-current protocols~\cite{MalcolmNicol2015,IllesCarbotteNicol2015,IllesNicol2016,IllesNicol2017,YeWangLai2024}. A key feature of the model is that one continuously tunable three-component spinor connects dispersive Dirac states with an exactly flat degree of freedom.

Throughout we write $\alpha=\tan\phi$ with hub--rim weights $c=\cos\phi$ and $u=\sin\phi$, so that the velocity scale $v_F$ is independent of $\alpha$ and the interpolation enters only through $(c,u)$; the intrinsic spin-orbit parameter $\lambda$ is the continuum energy multiplying the diagonal mass matrix rather than a bare next-nearest-neighbor hopping. With these conventions, the low-energy kinetic and spin-orbit structures coincide with established $\alpha$-$T_3$ continuum descriptions~\mbox{\cite{Raoux2014,WangLiu2021,TamangVermaBiswas2024,LeeEtAl2025}}, the dimensionless compensation ratios defined below are unique, and material-dependent calibration is treated separately from the continuum analysis.

The same three-sublattice structure also permits several distinct mass mechanisms to act on different components of the low-energy spinor. Intrinsic Kane--Mele spin--orbit coupling produces spin- and valley-dependent diagonal masses and gives an equilibrium quantum-spin-Hall transition at $\alpha=1/2$ in the lattice model~\cite{KaneMele2005QSH,KaneMele2005Z2,WangLiu2021}. The accompanying bulk geometry has been examined through orbital magnetization and circular dichroism~\cite{TamangVermaBiswas2024}, while edge-state analyses have emphasized that nonquantized Berry phase by itself neither determines nor excludes boundary modes in three-component Dirac systems~\cite{PratamaNakanishi2024}. Haldane and modified-Haldane extensions provide a complementary lattice route to Chern phases, indirect gap closings, topological metallicity, and edge spectra~\cite{LeeFuAng2024}. These results establish an important distinction for the present continuum problem: a local valley degeneracy is a spectral statement, whereas an integer topological invariant and the corresponding edge spectrum belong to a compact-Brillouin-zone formulation.

Periodic driving supplies a second, independently controllable mass. Off-resonant circular irradiation generates a sublattice-resolved mass through the leading high-frequency (van Vleck) commutator and has become a standard tool for synthesizing effective couplings and gauge structures~\mbox{\cite{GoldmanDalibard2014,EckardtAnisimovas2015,Bukov2015}}: circular light produces Hall-type Floquet phases in graphene and related lattices~\mbox{\cite{Kitagawa2011,OkaAoki2009,GomezLeon2014}}, Floquet--Bloch states have been observed spectroscopically in a driven topological insulator~\mbox{\cite{WangYH2013}}, and periodic protocols have been proposed to create and multiply Majorana modes in superconducting settings~\mbox{\cite{Jianglkt,TongQJ,WangZB}}. Controlled descriptions rest on quasienergy states and high-frequency expansions~\mbox{\cite{Sambe:1973cnm,Shirley:1965rgd,Magnus:1954zz,Blanes2009MagnusExpansion,Maricq1982hfeFloquetNmrOfSolids,Rahav2003HighFreqExp1}}, whose asymptotic character---and, where drive-induced absorption matters, finite prethermal window~\mbox{\cite{Prosen1999,Prosen2011,Ponte2015,Abanin2015,DAlessio2013,Mori2016}}---must be assessed rather than assumed. Within the $\alpha$-$T_3$ family, off-resonant light breaks valley and electron--hole symmetries, reshapes the flat band, redistributes Berry curvature, and induces band inversions~\mbox{\cite{DeyGhosh2018,DeyGhosh2019,Iurov2019,TamangBiswas2023}}. Because published low-energy conventions for the driven Hamiltonian differ---a discrepancy noted in a later Comment~\mbox{\cite{ChengComment2022}}---we derive the optical mass directly from minimal coupling and the leading van Vleck commutator, keeping valley $\eta$, real spin $s$, and helicity $\xi$ as independent labels. A full tight-binding analysis of the driven spin--orbit-coupled lattice has reported quantum-spin-Hall, Chern, and spin-polarized topological-metal regimes whose assignment depends on filling and on the participation of the middle band~\mbox{\cite{LeeEtAl2025}}; that global lattice setting complements the local, analytically controlled degeneracies studied here.

These two mass mechanisms thus act simultaneously on the same three-component spinor, and their competition poses the central question of this work: which pairwise band degeneracies emerge when the intrinsic and Floquet masses compensate, which of their properties are exact rather than perturbative, and can a charge transport observable not only detect but quantitatively invert this degeneracy structure?

\begin{figure}[t]
 \centering
 \includegraphics[width=\columnwidth]
 {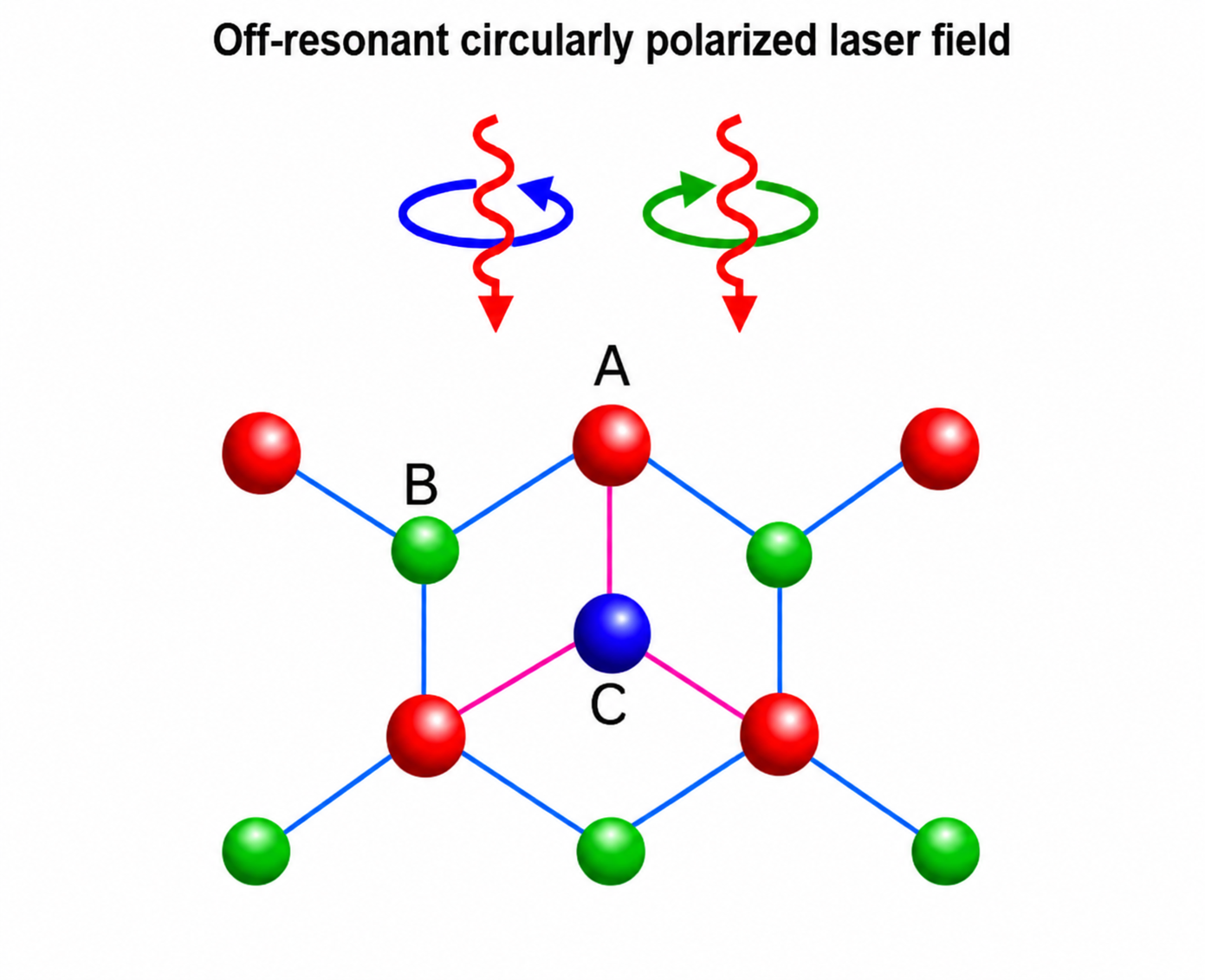}
 \caption{Schematic of the off-resonantly driven, spin--orbit-coupled $\alpha$-$T_3$ lattice. The three sublattices form the kinetic chain $A$--$B$--$C$ and are distinguished by color: $A$ (red), $B$ (green), and $C$ (blue); the sublattice-resolved spin--orbit and Floquet masses acting on the three on-site levels are indicated.}
 \label{fig:system_schematic}
\end{figure}

To answer this question, we combine the two diagonal mass mechanisms in the long-wavelength three-band Hamiltonian. Its kinetic connectivity is the chain $A$--$B$--$C$: the $A$--$B$ and $B$--$C$ matrix elements are linear in valley momentum, while the $A$ and $C$ components have no direct kinetic coupling. This elementary graph has a nontrivial consequence once spin-orbit and Floquet masses compensate. The three possible equalities of diagonal energies generate signed $AB$, $BC$, and $AC$ degeneracy lines. The first two are ordinary linear Dirac contacts because the degenerate states are connected directly. At the $AC$ condition, by contrast, the complete cubic characteristic polynomial factorizes exactly. One eigenstate becomes a dark rim-site combination with an energy independent of momentum throughout the continuum Hamiltonian, and the band that meets it disperses quadratically through virtual propagation across the nondegenerate $B$ state. The critical spectrum is thus an exact flat- and quadratic-band contact rather than a generic two-band quadratic crossing. Away from the contact, the corresponding two-band reduction carries winding-two and places the largest Berry curvature on a finite-momentum ring.

Keeping the optical amplitude and helicity as separate signed quantities gives an additional spin-selection rule for the degeneracy lines. The positive-drive $AB$ and $AC$ contacts reside in the spin sector opposite to the light helicity over the full interpolation interval. The $BC$ contact changes spin sector between $\alpha=1/2$ and $\alpha=1/\sqrt2$, reflecting the separate zeros of its spin--orbit and Floquet diagonal differences. The values $1/2$ and $1/\sqrt2$ already appear in equilibrium spin--orbit and irradiated $\alpha$-$T_3$ studies~\cite{WangLiu2021,DeyGhosh2019,Iurov2019,Cheng2020,Cheng2022}; here both follow from the signed compensation condition and directly identify the relevant real-spin sector. The full continuum Hamiltonian additionally obeys an exact opposite-valley conjugation relation, which constrains both its energy ordering and the chemical-potential parity of the valley-summed transverse responses.

Berry-curvature transport offers a way to probe this local spectral structure without assigning a global Chern number to the continuum theory. Anomalous Hall and transverse thermoelectric coefficients are particularly sensitive to the redistribution of Berry curvature near small gaps~\cite{XiaoYaoFangNiu2006,XiaoChangNiu2010}. Thermoelectric response in $\alpha$-$T_3$ systems has been investigated in magnetic-field, nanoribbon, optically driven, and spin--valley settings~\cite{Duan2023,YanWangLi2024,TamangBiswas2023,TamangBiswas2026}. Those calculations establish the relevance of the flat band, valley asymmetry, spin--orbit coupling, and Berry curvature to Seebeck and Nernst observables. Here we ask a more specific question: can the charge transverse thermoelectric response be inverted to recover parameters of the local three-band spectrum, rather than simply showing an extremum near a gap rearrangement?

For $0<\alpha<1/2$, the two linear degeneracy scales provide such an inverse relation. Their ratio is independent of the common spin--orbit energy scale and varies monotonically with $\alpha$. The response features associated with the linear $BC$ and $AB$ contacts can therefore determine $\alpha$ and the common spectral scale, after which the quadratic $AC$ spectral contact is predicted without using its thermoelectric extremum. The latter is displaced from the band contact by the momentum and entropy integrations. We test this construction with the complete three-band spectrum and Kubo response and determine the interval over which the inverse mapping remains identifiable.

The analysis yields three principal conclusions. First, the compensation of intrinsic and Floquet masses produces an exactly solvable degeneracy structure---most notably an exact flat- and quadratic-band contact protected by the absence of direct $A$--$C$ kinetic coupling---in a setting where driven three-band systems are usually treated only perturbatively. Second, the exact opposite-valley conjugation of the full Hamiltonian rigidly fixes the chemical-potential parity of the charge Hall and transverse thermoelectric responses, providing a symmetry check that survives any two-band reduction. Third, the scale-free ratio of the two linear degeneracy scales converts the transverse thermoelectric response into a quantitative parameter-inversion scheme whose domain of validity is explicitly delimited by the full three-band Kubo calculation. Together, these results tie the exact spectral algebra of a driven flat-band system to a measurable charge response and identify the observables that would test the construction experimentally.

The paper is organized as follows. Section~II derives the continuum Hamiltonian, the signed degeneracy loci, and the exact $AC$ factorization. Section~III introduces the Berry-curvature and thermoelectric formulation. Sections~IV and V discuss the critical spectrum and Berry-curvature structure. Sections~VI and VII develop the two-scale inversion and its range of validity, while Sec.~VIII states the numerical and physical regime of the calculation. Section~IX collects the main conclusions.

\section{Continuum Hamiltonian and exact degeneracy conditions}\label{sec:model}
The connectivity and sublattice-resolved spin--orbit and Floquet contributions of the driven $\alpha$-$T_3$ continuum model are summarized schematically in Fig.~\ref{fig:system_schematic}. The low-energy pristine $\alpha$-$T_3$ Hamiltonian follows from the graphene--dice interpolation introduced for the three-sublattice lattice~\cite{Raoux2014,MalcolmNicol2015,IllesNicol2016}. We expand the Hamiltonian about the two Dirac points and keep valley $\eta=\pm1$, real spin $s=\pm1$, and optical helicity $\xi=\pm1$ as distinct labels. The interpolation parameter is written as
\begin{align}
 \alpha&=\tan\phi,\qquad
 c\equiv\cos\phi=\frac{1}{\sqrt{1+\alpha^2}},\\
 u&\equiv\sin\phi=\frac{\alpha}{\sqrt{1+\alpha^2}}.
\end{align}
In the basis $(A,B,C)$ the low-energy kinetic Hamiltonian is
\begin{equation}
 H_0^\eta(\mathbf q)=
 \begin{pmatrix}
 0&c\gamma_\eta&0\\
 c\gamma_\eta^*&0&u\gamma_\eta\\
 0&u\gamma_\eta^*&0
 \end{pmatrix},\qquad
 \gamma_\eta=\hbar v_F(\eta q_x+i q_y).
 \label{eq:H0}
\end{equation}
The intrinsic Kane--Mele term is written in the continuum convention used for the spin-orbit-coupled $\alpha$-$T_3$ model~\cite{WangLiu2021,TamangVermaBiswas2024},
\begin{equation}
 H_{\rm so}^{\eta s}=\eta s\lambda\,
 \mathrm{diag}(-c,c-u,u).
 \label{eq:Hso}
\end{equation}
Here $\lambda$ is the continuum spin--orbit energy scale, while $s=\pm1$ and $\eta=\pm1$ label the real spin and valley, respectively. A comparison with microscopic next-nearest-neighbor hopping requires the same normalization convention for the velocity $v_F$ and $\alpha$ throughout. In the lattice $\alpha$-$T_3$ Kane--Mele model the intrinsic SOC is produced by next-nearest-neighbor hoppings of strength $\lambda$ on the $A$--$B$--$A$ paths and $\lambda_{KM}$ on the corresponding $C$--$B$--$C$ paths. The diagonal mass structure used in Eq.~\eqref{eq:Hso} is obtained by expanding the lattice Hamiltonian about a valley while keeping the leading momentum-independent SOC contribution. The resulting continuum SOC energy scale is denoted by $\lambda$ according to our convention, which keeps the overall Dirac velocity $v_F$ independent of $\alpha$. Therefore, $\lambda$ should be interpreted as the continuum Dirac mass produced by the microscopic Kane--Mele coupling rather than as the bare next-nearest-neighbor hopping itself. Specifically, the nearest-neighbor hopping convention and lattice spacing are crucial for the lattice-to-continuum conversion, whereas the corresponding $\alpha \lambda_{KM}$ hopping fixes the relative $\alpha$-dependence of the $C$-site SOC component.

For a circular vector potential
\begin{equation}
 \mathbf A(t)=A_0(\cos\omega t,\,\xi\sin\omega t),
\end{equation}
minimal coupling in Eq.~\eqref{eq:H0} gives Fourier components $H_{\pm1}$. Writing $a_0=e v_FA_0$, with $e>0$ the elementary-charge magnitude, they are
\begin{align}
H_{+1}=\frac{a_0}{2}
\begin{pmatrix}
0&c(\eta+\xi)&0\\
c(\eta-\xi)&0&u(\eta+\xi)\\
0&u(\eta-\xi)&0
\end{pmatrix},\nonumber\\
H_{-1}=\frac{a_0}{2}
\begin{pmatrix}
0&c(\eta-\xi)&0\\
c(\eta+\xi)&0&u(\eta-\xi)\\
0&u(\eta+\xi)&0
\end{pmatrix}.
\label{eq:Hpm}
\end{align}
For a monochromatic drive, the quasienergy construction of Floquet theory~\cite{Sambe:1973cnm,Shirley:1965rgd} reduces in the off-resonant regime to a hierarchy of virtual photon processes. Retaining the first nonvanishing contribution is the van Vleck form of the broader high-frequency framework~\cite{Magnus:1954zz,Rahav2003HighFreqExp1,GoldmanDalibard2014,EckardtAnisimovas2015,Bukov2015}; its use here is controlled by the scale hierarchy stated below. The leading van Vleck term in the standard high-frequency expansion~\cite{GoldmanDalibard2014,EckardtAnisimovas2015,Bukov2015} follows directly from the commutator,
\begin{align}
 \frac{[H_{-1},H_{+1}]}{\hbar\omega}
 &=\eta\xi\frac{a_0^2}{\hbar\omega}
 \mathrm{diag}(-c^2,c^2-u^2,u^2)\\
 &=\eta\xi\Delta_F\,
 \mathrm{diag}(-1,1-\alpha^2,\alpha^2),
 \label{eq:HF}
\end{align}
where
\begin{equation}
 \Delta_F=\frac{(e v_FA_0)^2}{(1+\alpha^2)\hbar\omega}.
 \label{eq:DeltaF}
\end{equation}
It is useful to state the high-frequency hierarchy in dimensionless form. With $r=\Delta_F/\lambda$ and $\Omega_\lambda=\hbar\omega/\lambda$,
\begin{equation}
 \epsilon_A\equiv\frac{e v_FA_0}{\hbar\omega}
 =\sqrt{\frac{r(1+\alpha^2)}{\Omega_\lambda}}.
 \label{eq:epsA}
\end{equation}
The leading van Vleck description applies when $\epsilon_A\ll1$ and the retained electronic scales, including $\lambda$, $\hbar v_Fq$, $|\mu|$, and $k_BT$, remain small compared with $\hbar\omega$. The sign and normalization of Eq.~\eqref{eq:HF} follow directly from the Fourier commutator in Eq.~\eqref{eq:Hpm}; in particular, reversing the helicity $\xi$ reverses the Floquet mass. 

The leading high-frequency Hamiltonian is
\begin{equation}
H_{\eta s}^{(\xi)}(\mathbf q)=
\begin{pmatrix}
m_A&c\gamma_\eta&0\\
c\gamma_\eta^*&m_B&u\gamma_\eta\\
0&u\gamma_\eta^*&m_C
\end{pmatrix},
\label{eq:Hfull}
\end{equation}
with
\begin{align}
 m_A&=\eta[-s\lambda c-\xi\Delta_F],\\
 m_B&=\eta[s\lambda(c-u)+\xi\Delta_F(1-\alpha^2)],\\
 m_C&=\eta[s\lambda u+\xi\Delta_F\alpha^2].
 \label{eq:masses}
\end{align}
The chain structure of Eq.~\eqref{eq:Hfull} gives the characteristic polynomial in closed form,
\begin{align}
 P(E,q)&\equiv\det[E\mathbb 1-H_{\eta s}^{(\xi)}(\mathbf q)]\nonumber\\
 &=(E-m_A)(E-m_B)(E-m_C)\nonumber\\
 &\quad-c^2|\gamma_\eta|^2(E-m_C)-u^2|\gamma_\eta|^2(E-m_A),
 \label{eq:charpoly}
\end{align}
where $|\gamma_\eta|^2=(\hbar v_F|q|)^2$. Equation~\eqref{eq:charpoly} establishes the radial spectral dependence directly and is used below to obtain an exact factorization at the $AC$ degeneracy.

Two exact relations follow directly from Eqs.~\eqref{eq:Hfull} and \eqref{eq:masses}. Since $\gamma_{-\eta}=-\gamma_{\eta}^{*}$ and every diagonal mass is odd in $\eta$,
\begin{equation}
 H_{-\eta,s}^{(\xi)}(\mathbf q)=-\left[H_{\eta,s}^{(\xi)}(\mathbf q)\right]^*.
 \label{eq:valley_conjugation}
\end{equation}
For energy-ordered bands $n=1,2,3$, Eq.~\eqref{eq:valley_conjugation} implies
\begin{align}
 E_{n,-\eta}(\mathbf q)&=-E_{4-n,\eta}(\mathbf q),\label{eq:valley_energy}\\
 \Omega_{n,-\eta}(\mathbf q)&=-\Omega_{4-n,\eta}(\mathbf q).\label{eq:valley_berry}
\end{align}
These identities determine the chemical-potential parity of the valley-summed charge responses discussed below. The Hamiltonian is traceless and rotationally invariant about each valley, which permits radial momentum integration in the continuum calculation.

\subsection{Signed pairwise degeneracy loci}\label{sec:criticalexact}
At $q=0$ all kinetic matrix elements vanish. It is then useful to introduce the signed optical variable
\begin{equation}
 \chi\equiv\xi\frac{\Delta_F}{\lambda},
 \label{eq:chi}
\end{equation}
which combines the two helicities on a single axis. With
\begin{equation}
 \mathbf g=(-c,c-u,u),\qquad \mathbf d=(-1,1-\alpha^2,\alpha^2),
\end{equation}
we have $m_i=\eta\lambda(sg_i+\chi d_i)$, and the condition $m_i=m_j$ gives
\begin{equation}
 \chi_{ij}=-s\frac{g_i-g_j}{d_i-d_j}.
 \label{eq:generalchi}
\end{equation}
Thus
\begin{align}
 \chi_{AB}&=-s\frac{2c-u}{2-\alpha^2}
 =-s\frac{2-\alpha}{(2-\alpha^2)\sqrt{1+\alpha^2}},\label{eq:chiAB}\\
 \chi_{BC}&=-s\frac{c-2u}{1-2\alpha^2}
 =-s\frac{1-2\alpha}{(1-2\alpha^2)\sqrt{1+\alpha^2}},\label{eq:chiBC}\\
 \chi_{AC}&=-s\frac{c+u}{1+\alpha^2}
 =-s\frac{1+\alpha}{(1+\alpha^2)^{3/2}}.
 \label{eq:chiAC}
\end{align}
These equations are exact within Eq.~\eqref{eq:Hfull}; no numerical root finder is used to obtain them.

For a positive drive amplitude at fixed helicity, $r\equiv\Delta_F/\lambda>0$ and $\chi=\xi r$. The $AB$ and $AC$ branches select $s=-\xi$ throughout $0\le\alpha\le1$. The numerator and denominator of Eq.~\eqref{eq:chiBC} change sign at different values, giving
\begin{equation}
 s_{BC}=\begin{cases}
 -\xi,&0\le\alpha<1/2,\\
 +\xi,&1/2<\alpha<1/\sqrt2,\\
 -\xi,&1/\sqrt2<\alpha\le1.
 \end{cases}
 \label{eq:spinBC}
\end{equation}
At $\alpha=1/2$ the $BC$ degeneracy occurs at $r=0$. At $\alpha=1/\sqrt2$ the $B$ and $C$ Floquet coefficients are equal while their SOI energies remain unequal, so the finite-$r$ solution is absent. At the two endpoints, all three pairwise values coalesce on the $s=-\xi$ branch,
\begin{align}
 r_{AB}=r_{BC}=r_{AC}&=1, && \alpha=0,\\
 r_{AB}=r_{BC}=r_{AC}&=\frac{1}{\sqrt2}, && \alpha=1.
\end{align}
For the numerical reference point $\alpha=0.30$ and $\xi=+1$,
\begin{align}
 r_{BC}&=0.4672323343, & r_{AB}&=0.8525155418,\\
 r_{AC}&=1.1423616246.&&
 \label{eq:reference_rc}
\end{align}
All three occur in the $s=-1$ sector. This common endpoint identifies the pseudospin-1 dice limit of the critical geometry, where the distinction between the three pairwise critical scales disappears. The corresponding Berry curvature reduces to the massive pseudospin-1 endpoint form derived and verified in Appendix~\mbox{\ref{app:endpointberry}}.

\subsection{Scale-free inversion from the two linear branches}\label{sec:ratioinversion}
For $0<\alpha<1/2$ and fixed helicity, the positive-drive $BC$ and $AB$ branches belong to the same real-spin sector and both are linear Dirac touchings. Their ratio eliminates any common energy calibration,
\begin{equation}
 \mathcal R_{BA}(\alpha)\equiv\frac{r_{BC}}{r_{AB}}
 =\frac{(1-2\alpha)(2-\alpha^2)}{(1-2\alpha^2)(2-\alpha)}.
 \label{eq:RBA}
\end{equation}
This ratio is strictly monotonic and therefore one-to-one on the interval. Direct differentiation gives
\begin{align}
 \frac{d\mathcal R_{BA}}{d\alpha}
 &=-\frac{3\left[2\alpha^2(\alpha-1)^2+3\alpha^2-4\alpha+2\right]}
 {(2-\alpha)^2(1-2\alpha^2)^2},\\
 \frac{d\mathcal R_{BA}}{d\alpha}&<0,
 \qquad 0<\alpha<\frac12.
 \label{eq:RBAmonotonic}
\end{align}
The bracket in the numerator is strictly positive because $3\alpha^2-4\alpha+2>0$. Moreover $\mathcal R_{BA}(0)=1$ and $\mathcal R_{BA}(1/2)=0$. A measured ratio of the two linear critical drives then determines a unique $\alpha$ in this interval without knowledge of the common energy scale.

Let $D$ denote an experimental drive coordinate proportional to the effective Floquet scale for a fixed sample. At a fixed frequency the drive coordinate may be taken as the optical intensity, $D\propto A_0^2$, which for a calibrated frequency protocol is equivalent to $A_0^2/\omega$ up to a constant. Since at fixed $\alpha$ the Floquet scale obeys $r\propto A_0^2/\omega$, all three critical values in one sample satisfy $D_i=\Lambda r_i(\alpha)$ with a common multiplicative factor $\Lambda$. The field amplitude $A_0$ itself is not the linear coordinate entering the ratio. After solving Eq.~\eqref{eq:RBA} from $D_{BC}/D_{AB}$, the two linear features determine $\Lambda$ by least squares. The quadratic spectral crossing is then predicted without using its thermoelectric extremum,
\begin{equation}
 D_{AC}^{\rm pred}=\Lambda\,
 \frac{1+\alpha}{(1+\alpha^2)^{3/2}}.
 \label{eq:ACprediction}
\end{equation}
This overdetermination is used below as a test of the continuum model: the $BC$ and $AB$ response features fix $(\alpha,\Lambda)$, while $AC$ is held out from the parameter fit.

The pairwise crossings are isolated for $0<\alpha<1$ except at the stated coalescence/singular limits. Substitution of the signed critical drives into the diagonal masses gives the three crossing energies,
\begin{align}
 E_c^{AB}&=-\frac{\eta s\lambda\alpha(1-\alpha)}{(2-\alpha^2)\sqrt{1+\alpha^2}},\\
 E_c^{BC}&=\frac{\eta s\lambda\alpha(1-\alpha)}{(1-2\alpha^2)\sqrt{1+\alpha^2}},\\
 E_c^{AC}&=\frac{\eta s\lambda\alpha(1-\alpha)}{(1+\alpha^2)^{3/2}}.
 \label{eq:crossing_energies}
\end{align}
These are the repeated $q=0$ eigenvalues of the corresponding pairwise degeneracies. The separation between $E_c$ and the remaining sublattice level is
\begin{align}
E_c-E_C\big|_{AB}&=-\frac{3\eta s\lambda\alpha(1-\alpha)}{(2-\alpha^2)\sqrt{1+\alpha^2}},\\
E_c-E_A\big|_{BC}&=\frac{3\eta s\lambda\alpha(1-\alpha)}{(1-2\alpha^2)\sqrt{1+\alpha^2}},\\
E_c-E_B\big|_{AC}&=\frac{3\eta s\lambda\alpha(1-\alpha)}{(1+\alpha^2)^{3/2}}.
\label{eq:remote_gaps}
\end{align}
The $AC$ result is finite for every $0<\alpha<1$, which justifies a two-state Schrieffer--Wolff reduction at an interior $AC$ crossing.

\subsection{Linear $AB/BC$ and quadratic $AC$ touchings}\label{sec:effective}
Let
\begin{equation}
 M_{ij}=\frac{m_i-m_j}{2}
 =\frac{\eta\xi\lambda}{2}(d_i-d_j)(r-r_{ij}).
 \label{eq:Mij}
\end{equation}
For the directly connected pairs, projection onto the degenerate states gives
\begin{align}
 H_{AB}^{\rm eff}&=E_c\mathbb 1+
 \begin{pmatrix}M_{AB}&c\gamma_\eta\\c\gamma_\eta^*&-M_{AB}\end{pmatrix}+O(q^2),\label{eq:HeffAB}\\
 H_{BC}^{\rm eff}&=E_c\mathbb 1+
 \begin{pmatrix}M_{BC}&u\gamma_\eta\\u\gamma_\eta^*&-M_{BC}\end{pmatrix}+O(q^2).
 \label{eq:HeffBC}
\end{align}
At $M_{ij}=0$, their direct gaps are proportional to $q$. More generally, the kinetic matrix has the connectivity graph $A$--$B$--$C$. Each kinetic edge is linear in momentum, so the leading coupling between two degenerate basis states occurs at the order set by the shortest path connecting them, provided all intermediate states remain nondegenerate. The direct pairs have path length one; the $A$--$C$ pair has path length two.

There is no direct $A$--$C$ element in Eq.~\eqref{eq:Hfull}. At the $AC$ degeneracy, $m_A=m_C\equiv E_c$, while the $B$ level is separated by
\begin{equation}
 \Delta_B^{AC}\equiv E_c-m_B
 =\frac{3\eta s\lambda\,\alpha(1-\alpha)}{(1+\alpha^2)^{3/2}}.
 \label{eq:remoteAC}
\end{equation}
The separation is nonzero for every interior $0<\alpha<1$ and vanishes at the endpoint triple degeneracies. More can be obtained here than a perturbative statement. Substitution of $m_A=m_C=E_c$ into Eq.~\eqref{eq:charpoly} gives the exact factorization
\begin{equation}
 P(E,q)=(E-E_c)\left[(E-E_c)(E-m_B)-|\gamma_\eta|^2\right].
 \label{eq:ACfactor}
\end{equation}

One eigenvalue is consequently independent of momentum,
\begin{equation}
 E_{\rm flat}(q)=E_c,
 \label{eq:ACflat}
\end{equation}
which is an exact dark-state flat band generated by destructive interference, and the two remaining eigenvalues are
\begin{equation}
 E_{\pm}(q)=E_c+\frac{-\Delta_B\pm\sqrt{\Delta_B^2+4|\gamma_\eta|^2}}{2}.
 \label{eq:ACexactspectrum}
\end{equation}
The root that meets the flat band at $q=0$ can be written as
\begin{align}
 E_{\rm q}(q)-E_c
 &=\frac{-\Delta_B+\operatorname{sgn}(\Delta_B)
 \sqrt{\Delta_B^2+4|\gamma_\eta|^2}}{2}\nonumber\\
 &=\frac{|\gamma_\eta|^2}{\Delta_B}
 -\frac{|\gamma_\eta|^4}{\Delta_B^3}+O(q^6).
 \label{eq:ACquadraticexact}
\end{align}
Thus the $AC$ critical point is an exact flat- and quadratic-band contact within the continuum Hamiltonian. For $q\neq0$, the flat eigenvalue is carried by the dark state
\begin{equation}
 |D_\eta(\mathbf q)\rangle=
 \begin{pmatrix}
 u\,\gamma_\eta/|\gamma_\eta|\\[2pt]
 0\\[2pt]
 -c\,\gamma_\eta^*/|\gamma_\eta|
 \end{pmatrix},
 \qquad H|D_\eta\rangle=E_c|D_\eta\rangle .
 \label{eq:ACdark}
\end{equation}
The state has zero $B$-sublattice amplitude because the two kinetic paths into $B$ interfere destructively. This result is structural: for any three-site chain with equal endpoint energies and couplings proportional to the same complex momentum amplitude, the linear combination orthogonal to the hub coupling is an exact eigenstate. The optical and spin--orbit terms in Eq.~\eqref{eq:Hfull} restore this condition precisely on the $AC$ degeneracy line.

The dark state also carries a closed-loop geometric phase that follows directly from Eq.~\eqref{eq:ACdark}. On a circle $\mathbf q=q(\cos\theta,\sin\theta)$,
\begin{equation}
 i\langle D_\eta|\partial_\theta D_\eta\rangle
 =\eta(c^2-u^2),
\end{equation}
so that
\begin{equation}
 \gamma_D=2\pi\eta\frac{1-\alpha^2}{1+\alpha^2}\pmod{2\pi}.
 \label{eq:darkBerryPhase}
\end{equation}
This loop phase is defined away from the degeneracy at $q=0$; it is not a Brillouin-zone Chern number.

The Schrieffer--Wolff reduction provides the detuned two-band description that controls the Berry curvature. Let $P=|A\rangle\langle A|+|C\rangle\langle C|$, $Q=|B\rangle\langle B|$, and $M_{AC}=(m_A-m_C)/2$. To second order in momentum,
\begin{equation}
H_{AC}^{\rm eff}=E_c\mathbb 1+
\begin{pmatrix}
M_{AC}+A_q&V_q\\
V_q^*&-M_{AC}+C_q
\end{pmatrix},
\label{eq:HeffAC}
\end{equation}
where
\begin{equation}
 A_q=\frac{c^2|\gamma|^2}{\Delta_B},\qquad
 C_q=\frac{u^2|\gamma|^2}{\Delta_B},\qquad
 V_q=\frac{cu\gamma_\eta^2}{\Delta_B}.
 \label{eq:ACmatrixelements}
\end{equation}
At the critical line the next correction to the dispersive eigenvalue is $O(q^4)$, as seen explicitly in Eq.~\eqref{eq:ACquadraticexact}; away from the line, the coefficients acquire corrections of order $M_{AC}q^2$. The off-diagonal phase is $2\eta\theta_{\mathbf q}$ and winds by $4\pi$ around a valley.
The winding-two structure of Eq.~\eqref{eq:HeffAC} describes the local pseudospin texture of the $AC$ effective Hamiltonian; identifying it with an integer Chern number requires a globally regularized Brillouin-zone Hamiltonian. Winding-two quadratic Hamiltonians also arise in chiral bilayer graphene and in Floquet systems with quadratic band-touching points~\cite{McCannFalco2006,DuZhouFiete2017}; here the structure follows from the $A$--$B$--$C$ kinetic connectivity and the exact flat eigenstate in Eq.~\eqref{eq:ACdark}.

The low-energy density of states reflects the distinction between the three crossings. A linear $AB$ or $BC$ cone gives $\rho(E)\propto|E-E_c|$. At the exact $AC$ line, a finite momentum patch contains a cutoff-dependent flat-band weight $W_{\rm flat}\delta(E-E_c)$ together with a one-sided constant contribution from the quadratic dispersive branch,
\begin{equation}
 \rho_{AC}(E)=W_{\rm flat}\delta(E-E_c)
 +C_{AC}\,\Theta[(E-E_c)\Delta_B]+\cdots .
 \label{eq:DOS_AC}
\end{equation}
This singular flat-band term is absent from a generic two-band quadratic touching and is retained in the full three-band numerical spectrum.

Writing the traceless part of Eq.~\eqref{eq:HeffAC} as $\mathbf d\cdot\boldsymbol\sigma$ gives
\begin{align}
 d_x&=\beta(q_x^2-q_y^2),\qquad d_y=-2\eta\beta q_xq_y,\\
 d_z&=M_{AC}+Bq^2,
\end{align}
with $\beta=cu(\hbar v_F)^2/\Delta_B$ and $B=(c^2-u^2)(\hbar v_F)^2/(2\Delta_B)$. Its two-band Berry curvature is
\begin{equation}
 \Omega_{\pm}^{AC}(q)=\pm\frac{2\eta\beta^2 M_{AC}q^2}
 {\left[\beta^2q^4+(M_{AC}+Bq^2)^2\right]^{3/2}}.
 \label{eq:OmegaACeff}
\end{equation}
Unlike the massive linear Dirac curvature, Eq.~\eqref{eq:OmegaACeff} vanishes at $q=0$ for nonzero $M_{AC}$ and reaches its maximum on a finite-$q$ ring. The radius can be obtained analytically. With $A_2=\beta^2+B^2$ and $s_M=\operatorname{sgn}M_{AC}$, maximizing $|\Omega^{AC}(q)|$ gives
\begin{align}
 q_{\Omega}^2&=|M_{AC}|\,y_*,\nonumber\\
 y_*&=\frac{-s_M B+\sqrt{B^2+8A_2}}{4A_2}.
 \label{eq:qOmega_exact}
\end{align}
Consequently,
\begin{equation}
 q_{\Omega}\propto |M_{AC}|^{1/2}\propto |r-r_{AC}|^{1/2}.
 \label{eq:qOmega_scaling}
\end{equation}
Figure~\ref{fig:berry}(c) compares the maximum of the full three-band Kubo curvature with Eq.~\eqref{eq:qOmega_exact}, whose coefficients are fixed by the Schrieffer-Wolff Hamiltonian.

\section{Berry curvature and thermoelectric formulation}\label{sec:numerics}
For a nondegenerate eigenstate of Eq.~\eqref{eq:Hfull}, the Berry curvature is evaluated from the interband Kubo formula~\cite{XiaoChangNiu2010},
\begin{equation}
 \Omega_n(\mathbf q)=-2\,\mathrm{Im}\sum_{m\ne n}
 \frac{\langle n|\partial_{q_x}H|m\rangle
 \langle m|\partial_{q_y}H|n\rangle}
 {(E_n-E_m)^2}.
 \label{eq:KuboBerry}
\end{equation}
This expression avoids derivatives of the eigenvector phase. Individual-band curvature is not defined at an exact degeneracy, so all band-resolved evaluations use a nonzero momentum or parameter offset. The complete three-band sum satisfies $\sum_n\Omega_n=0$ away from degeneracies.

With $e>0$ denoting the elementary-charge magnitude, the intrinsic Hall and transverse thermoelectric conductivities are~\cite{XiaoYaoFangNiu2006,XiaoChangNiu2010}
\begin{align}
 \sigma_{xy}^{\eta s}&=-\frac{e^2}{\hbar}\sum_n\int\frac{d^2q}{(2\pi)^2}
 f_{n\eta s}\,\Omega_{n\eta s},\label{eq:sigma}\\
 \alpha_{xy}^{\eta s}&=\frac{e k_B}{\hbar}\sum_n\int\frac{d^2q}{(2\pi)^2}
 \mathcal S(f_{n\eta s})\,\Omega_{n\eta s},\label{eq:alphaN}
\end{align}
where $\mathcal S(f)=-f\ln f-(1-f)\ln(1-f)$. Rotational symmetry reduces the momentum integral to a radial quadrature. Charge, spin, valley, and spin--valley combinations follow from
\begin{equation}
 X^{c,s,v,sv}=\sum_{\eta,s}(1,s,\eta,\eta s)X^{\eta s}.
 \label{eq:channels}
\end{equation}
The inverse analysis uses the charge coefficient $\alpha_{xy}^{c}$. This quantity is the transverse thermoelectric conductivity, often called the anomalous Nernst conductivity. The open-circuit Nernst thermopower additionally contains longitudinal transport coefficients, so it is a distinct observable.

Equation~\eqref{eq:valley_conjugation} also constrains the transport integrals. Using $\mathcal S[-E-\mu]=\mathcal S[E+\mu]$ after the valley and band-order transformation gives
\begin{align}
 \sigma_{xy}^{c}(\mu)&=\sigma_{xy}^{c}(-\mu),\label{eq:hall_even}\\
 \alpha_{xy}^{c}(\mu)&=-\alpha_{xy}^{c}(-\mu).\label{eq:nernst_odd}
\end{align}
At low temperature, the entropy integral approaches the Mott relation for the intrinsic contribution~\cite{XiaoYaoFangNiu2006},
\begin{equation}
 \alpha_{xy}^{c}\simeq-\frac{\pi^2 k_B^2T}{3e}
 \frac{\partial\sigma_{xy}^{c}}{\partial\mu}.
 \label{eq:Mott}
\end{equation}
The occupation factors are taken in the effective-equilibrium form appropriate to an off-resonant prethermal or bath-stabilized regime, where $H_{\rm eff}$ is described by effective parameters $T$ and $\mu$~\cite{HoMoriAbaninDallaTorre2023}. The transport coefficients below are intrinsic coefficients of this leading high-frequency Hamiltonian.

For a response array $A(r_i,\mu_j)=\alpha_{xy}^{c}(r_i,\mu_j)$, drive-localized changes are summarized by
\begin{equation}
 \mathcal Q(r_i)=\left[\frac{1}{N_\mu}\sum_j
 \left(\frac{\partial A(r_i,\mu_j)}{\partial r}\right)^2\right]^{1/2}.
 \label{eq:Q}
\end{equation}

We do not propose $Q(r)$ as a directly measured observable; rather, we introduce it as a model-independent numerical feature detector for identifying drive-dependent features in the computed response. Response scales are located as persistent maxima of $\mathcal Q(r)$ using one fixed multiscale prominence criterion throughout the parameter scans. In the interval $0<\alpha<1/2$, the two lowest resolved structures are assigned to the $BC$ and $AB$ branches according to the analytic ordering in Eq.~\eqref{eq:reference_rc}; the $AC$ structure is then used as a prediction test rather than as an input to the two-scale inversion.

In the interval $0<\alpha<1/2$, uncertainty in the scale-free two-linear-feature inversion follows directly from Eq.~\eqref{eq:RBA}. If $\mathbf D=(D_{BC},D_{AB})$ has covariance matrix $C$ and $R=D_{BC}/D_{AB}$, then
\begin{align}
 \nabla_{\!D}R&=\left(\frac{1}{D_{AB}},-\frac{D_{BC}}{D_{AB}^2}\right),\\
 \sigma_\alpha^2&=\frac{(\nabla_{\!D}R)^T C(\nabla_{\!D}R)}
 {[\mathcal R_{BA}'(\alpha)]^2}.
 \label{eq:errorprop}
\end{align}
Here $C$ is the covariance matrix of the extracted feature positions
$\mathbf D=(D_{BC},D_{AB})$, and Eq.~\mbox{\eqref{eq:errorprop}} follows from
the first-order relations $\delta R=(\nabla_D R)\cdot\delta\mathbf D$
and $\delta\alpha=\delta R/\mathcal R'_{BA}(\alpha)$. For the reference
numerical calculation below, a $1\%$ additive perturbation of the
response gives $\sigma_\alpha\simeq1.5\times10^{-4}$.

\begin{figure*}[t]
\includegraphics[width=0.98\textwidth]{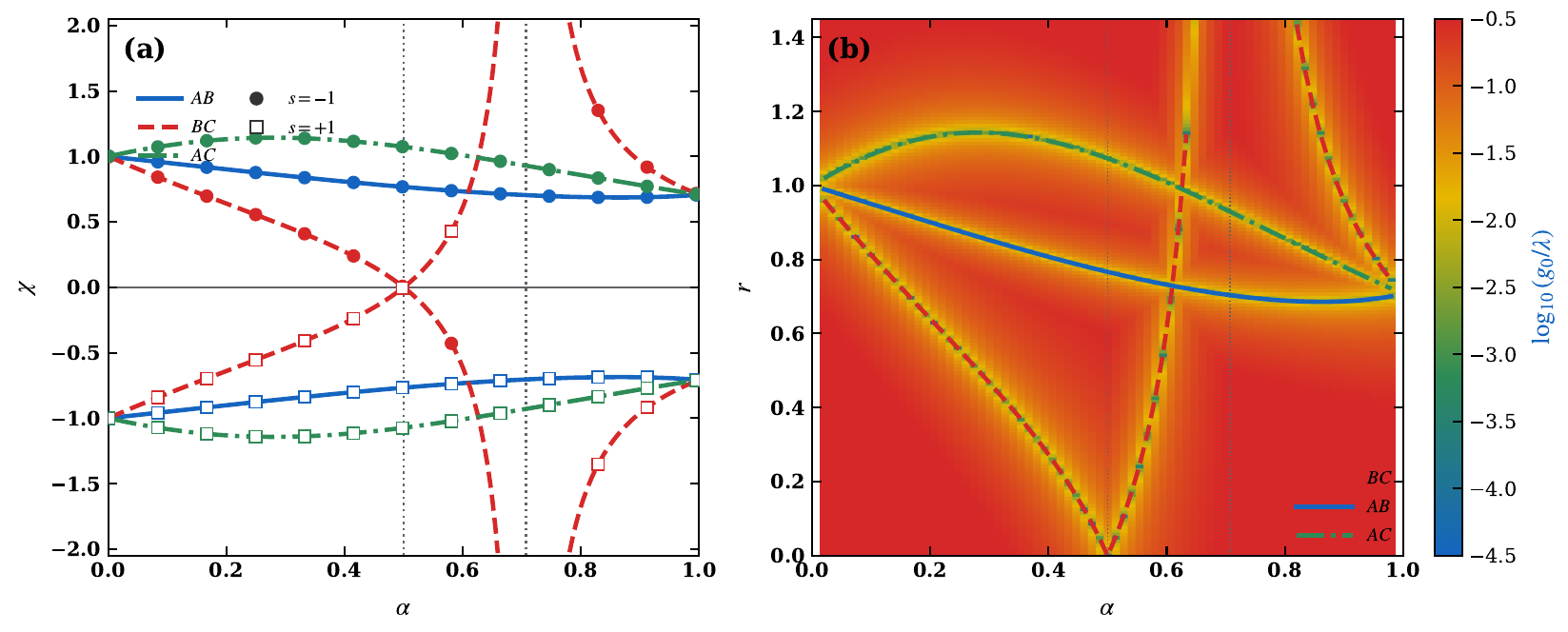}
\caption{Signed and positive-drive critical geometry. (a) Exact pairwise degeneracies $\chi_{ij}$. The three branches are distinguished by both color and line style: $AB$ is shown by blue solid lines, $BC$ by orange dashed lines, and $AC$ by magenta dash-dotted lines. The real-spin sectors are further distinguished by filled-circle markers for $s=-1$ and open-square markers for $s=+1$. The dotted vertical lines mark $\alpha=1/2$ and $1/\sqrt{2}$. The $BC$ branch passes through zero at $\alpha=1/2$ and has no finite-drive degeneracy at $\alpha=1/\sqrt2$, where the $B$ and $C$ Floquet coefficients become equal. (b) Minimum adjacent $q=0$ gap $g_0$ obtained from direct diagonalization of the complete three-band Hamiltonian and minimization over real spin at fixed $\xi=+1$. The analytic positive-drive degeneracies are superposed on the gap map using the same branch-specific colors and line styles as in panel (a).}
\label{fig:critical}
\end{figure*}

\begin{figure}[t]
 \centering
 \includegraphics[width=\columnwidth]{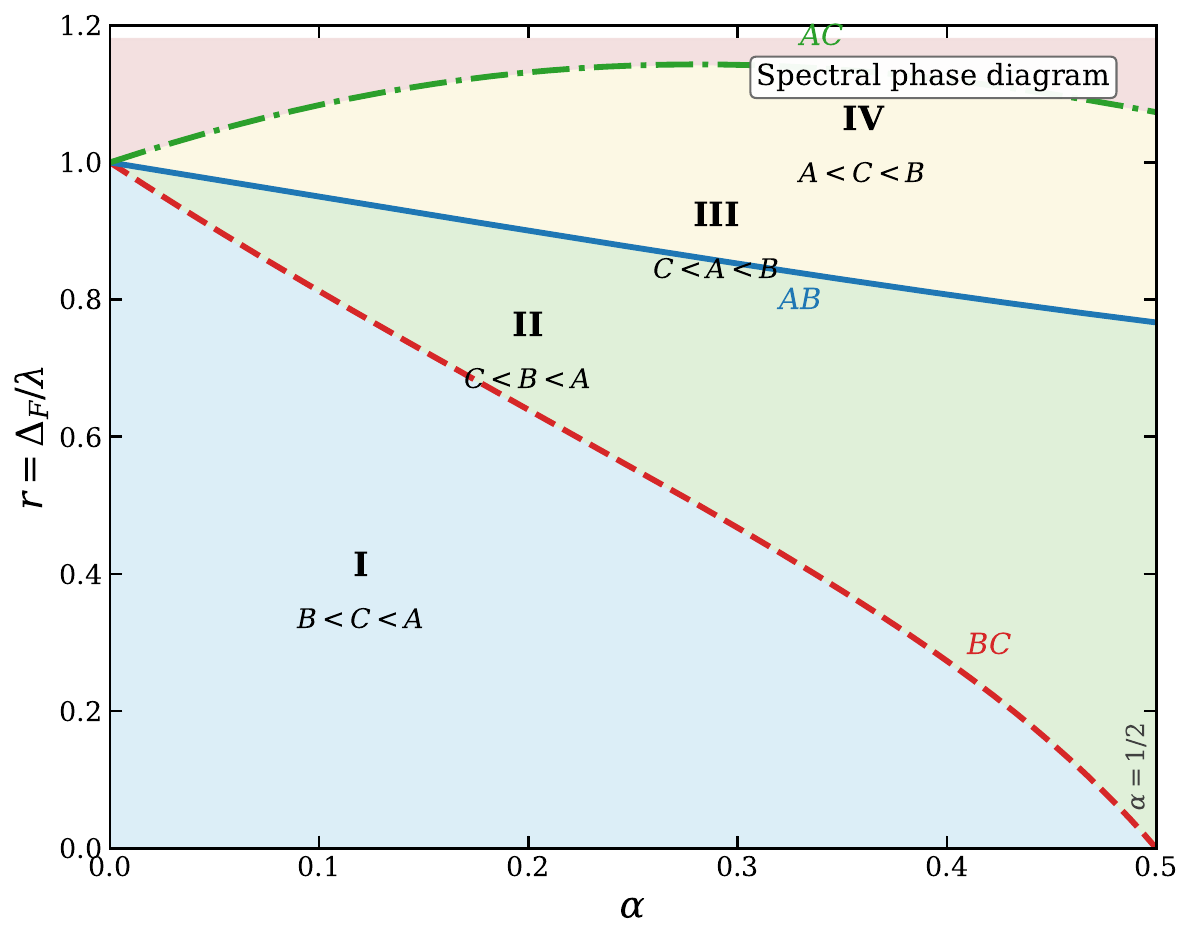}
 \caption{Spectral phase diagram of the positive-drive branch in the interval $0\le \alpha \le 1/2$. The exact compensation lines $BC$, $AB$, and $AC$ divide the $(\alpha,r)$ plane, with $r=\Delta_F/\lambda$, into four regions with distinct $q=0$ level orderings. As the drive increases at fixed $\alpha$, the sequence of spectral rearrangements is $BC\rightarrow AB\rightarrow AC$. The $BC$ and $AB$ boundaries correspond to linear Dirac contacts, whereas the $AC$ boundary marks the restoration of the exact dark flat band and its quadratic touching partner. At the graphene endpoint $\alpha=0$ the three boundaries emanate from the single point $(\alpha,r)=(0,1)$, where the two intermediate orderings collapse because the $C$ level stays at zero energy for every drive. The dice endpoint $\alpha=1$ lies outside the plotted interval and is shown in Fig.~\mbox{\ref{fig:critical}}(a).}
 \label{fig:phase_diagram}
\end{figure}

\section{Spectral critical geometry}\label{sec:spectral_results}
Throughout the numerical calculations we use dimensionless units with $\lambda=1$ and $\hbar v_F=1$ (energies in units of $\lambda$ and momenta in units of $\lambda/\hbar v_F$); taking $\hbar\omega/\lambda=100$ for the high-frequency consistency check, the largest reference critical drive at $\alpha=0.30$, $r_{AC}=1.14236$, gives $\epsilon_A=\sqrt{r(1+\alpha^2)/(\hbar\omega/\lambda)}\simeq0.112\ll1$, confirming that the required drive lies within the leading van Vleck regime. Figure~\ref{fig:critical} compares the analytic pairwise degeneracies with a direct diagonalization of the complete Hamiltonian at $q=0$. Panel (a) uses the signed coordinate $\chi=\xi\Delta_F/\lambda$ and displays both real-spin sectors. The $AB$ and $AC$ branches remain finite across the interpolation range. The $BC$ branch passes through zero at $\alpha=1/2$ and has no finite-drive degeneracy at $\alpha=1/\sqrt2$, where the $B$ and $C$ Floquet coefficients become equal. Panel (b) is obtained by diagonalizing Eq.~\eqref{eq:Hfull} for both spins at every grid point and taking the smallest adjacent gap. The analytic curves are superposed on the resulting gap map. Their coincidence with the numerical minima confirms Eqs.~\eqref{eq:chiAB}--\eqref{eq:chiAC}.

The same critical lines partition the positive-drive $(\alpha,r)$ plane into four spectral regions, as summarized in Fig.~\mbox{\ref{fig:phase_diagram}}; in particular, the $AC$ branch $r_{AC}(\alpha)$ forms the boundary between regions III and IV and marks the transition through the exact flat--quadratic contact. For $0<\alpha<1/2$, increasing $r$ successively crosses the $BC$, $AB$, and $AC$ boundaries, changing the $q=0$ level ordering from $B<C<A$ to $C<B<A$, $C<A<B$, and finally $A<C<B$. The diagram is restricted to the inversion sector $0<\alpha<1/2$; beyond this interval the $BC$ branch changes spin sector, while at $\alpha=1/\sqrt{2}$ no finite-drive $BC$ compensation exists because the $B$ and $C$ Floquet coefficients become equal. The restricted phase diagram in Fig.~\mbox{\ref{fig:phase_diagram}} isolates the $0<\alpha<1/2$ sector in which the two-linear-feature inversion is uniquely defined, whereas Fig.~\mbox{\ref{fig:critical}} spans the full interpolation range. The two endpoints follow directly from the diagonal masses and appear as the degenerate limits of this partition. At $\alpha=0$, $u=0$ and $\alpha^2=0$ pin the $C$ level to $m_C=0$ for every drive while $m_A=-m_B$: the three boundary curves meet at the triple point $(\alpha,r)=(0,1)$ in Fig.~\mbox{\ref{fig:phase_diagram}}, where $m_A=m_B=m_C=0$ and the two intermediate orderings collapse, leaving only $B<C<A$ below and $A<C<B$ above $r=1$. At $\alpha=1$, symmetrically, $c=u$ and $1-\alpha^2=0$ pin the hub level to $m_B=0$, and the three pairwise conditions coalesce at $r=1/\sqrt{2}$ into the same threefold $q=0$ degeneracy, with only $C<B<A$ and $A<B<C$ on either side; this endpoint lies outside the restricted interval of the phase diagram and is read off Fig.~\mbox{\ref{fig:critical}}(a).

\begin{figure*}[t]
\includegraphics[width=0.98\textwidth]{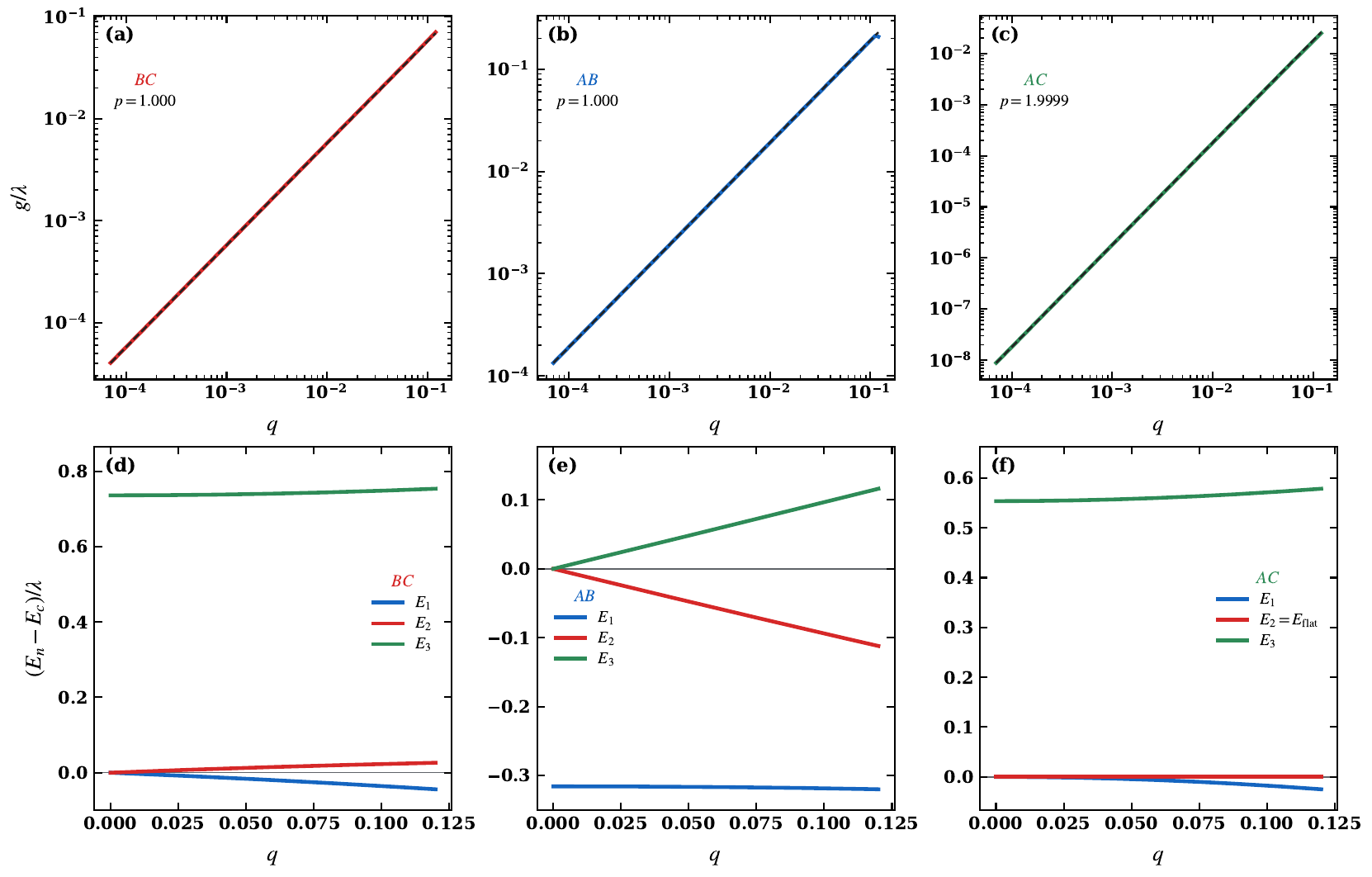}
\caption{Critical dispersion from direct diagonalization of the complete $3\times3$ Hamiltonian at $\alpha=0.30$. (a)--(c) Minimum adjacent gap $g$ and small-$q$ power-law fits: linear $g\propto q$ for the $BC$ and $AB$ contacts, quadratic $g\propto q^2$ at $AC$. (d)--(f) The same full calculation plotted as $(E_n-E_c)/\lambda$, with one common crossing energy $E_c$ in each panel. All three energy-ordered eigenvalues are shown and annotated. The remote band remains separated from the touching pair; no two-band eigenvalue is substituted into these panels. At $AC$, $E_2=E_{\rm flat}=E_c$ is the exact dark flat band, $E_1$ is its quadratic partner, and $E_3$ remains nondegenerate.}
\label{fig:dispersion}
\end{figure*}

The local momentum dependence separates the three degeneracies into two classes. Figure~\ref{fig:dispersion} shows the spectrum obtained by diagonalizing the full Hamiltonian at the exact critical drives. The $BC$ and $AB$ adjacent gaps are linear in $q$, in agreement with the direct matrix elements in Eqs.~\eqref{eq:HeffAB} and \eqref{eq:HeffBC}. At the $AC$ point, the minimum gap is quadratic. Panels (d)--(f) retain all three energy-ordered eigenvalues of the complete Hamiltonian and plot them relative to the single crossing energy $E_c$. The remote eigenvalue is kept on the same ordinate rather than removed by a low-energy projection. At the chosen reference point, the remote level lies $0.736\,\lambda$ from the $BC$ crossing, $0.316\,\lambda$ from the $AB$ crossing, and $0.554\,\lambda$ from the $AC$ crossing. Panel (f) displays the stronger exact statement from Eq.~\eqref{eq:ACfactor}: $E_2=E_{\rm flat}=E_c$ remains dispersionless throughout the plotted continuum interval, while $E_1$ departs quadratically and $E_3$ remains separated. Appendix~\ref{app:bandshifts} provides a complementary bandwise representation that isolates the momentum-induced dispersion of each band.

The quadratic law follows from the shortest kinetic path in the sublattice graph $A$--$B$--$C$. Directly connected pairs acquire a first-order matrix element proportional to $q$, whereas the $A$ and $C$ states communicate through two kinetic matrix elements and the finite denominator $\Delta_B^{AC}$. More strongly, Eq.~\eqref{eq:ACfactor} factorizes the complete characteristic polynomial: one eigenvalue remains exactly flat, and the branch that meets it has the expansion in Eq.~\eqref{eq:ACquadraticexact}, whose first correction is $O(q^4)$.

The singular density of states at the $AC$ line also differs from the usual two-band quadratic crossing. A low-energy momentum patch contains a delta-function contribution from the exact flat band and a constant one-sided contribution from the quadratic branch, Eq.~\eqref{eq:DOS_AC}. The linear $AB$ and $BC$ points instead give the familiar two-dimensional Dirac law $\rho(E)\propto|E-E_c|$. Figure~\ref{fig:Sphase} in Appendix~\ref{app:acfactor} evaluates the local spectral weight directly from the three numerical eigenvalues and recovers the cumulative phase-space powers $\mathcal N_{\rm disp}\propto\varepsilon^2$ for $AB/BC$ and $\mathcal N_{\rm disp}\propto\varepsilon$ for the dispersive $AC$ branch. Such flat- and quadratic-band contact points have appeared in other Floquet multiband settings~\mbox{\cite{DuZhouFiete2017}}; here both arise from the same three-sublattice continuum Hamiltonian and can be related algebraically to its chain connectivity.

The dark eigenstate itself retains the continuously tunable geometric structure of the $\alpha$-$T_3$ spinor. Equation~\eqref{eq:darkBerryPhase} gives the loop Berry phase $\gamma_D=2\pi\eta(1-\alpha^2)/(1+\alpha^2)$ modulo $2\pi$. The phase evolves continuously between the graphene and dice endpoints and follows directly from the relative rim-site amplitudes of the exact dark state. This geometric phase characterizes the local eigenvector winding of the critical flat band; the global topological classification remains a lattice property.

Off-resonant irradiation generally deforms the flat band of the $\alpha$-$T_3$ spectrum away from the graphene and dice limits~\cite{DeyGhosh2018,TamangBiswas2023}. The flat eigenvalue found here is more restrictive: it is restored exactly on the spin--orbit--Floquet compensation line $m_A=m_C$. The condition is a codimension-one spectral constraint of the combined problem, rather than generic persistence of the undriven flat band. In the full lattice model, the same drive and spin--orbit terms also enter the Chern and spin-Chern phase structure~\cite{LeeEtAl2025}; the factorization above singles out the local three-band physics of the rim-site degeneracy.

The special interpolation values connect naturally with established lattice results. At $\alpha=1/2$, the intrinsic $B$--$C$ spin--orbit splitting vanishes and the corresponding degeneracy sits at zero optical drive, the same value at which the equilibrium spin--orbit lattice problem changes its quantum-spin-Hall sector~\cite{WangLiu2021}. At $\alpha=1/\sqrt2$, the leading Floquet shifts of $B$ and $C$ are identical, so varying the leading optical mass cannot remove their residual spin--orbit splitting. Spinless irradiated models also single out this interpolation value~\cite{DeyGhosh2019,Iurov2019}. The recent lattice spin--orbit--Floquet phase diagram contains the associated global topological information~\cite{LeeEtAl2025}; the equations above describe the local continuum degeneracies that enter that broader setting.

At the reference value used in Figs.~\ref{fig:dispersion}--\ref{fig:inverse}, the $BC$ and $AC$ conditions close the lower adjacent gap, whereas $AB$ closes the upper gap. Scanning the chemical potential probes the relevant spectral neighborhoods and can retain signatures of all three contacts. At fixed insulating filling, only the adjacent gap belonging to the occupied manifold is relevant. This filling dependence is essential when local gap zeros are compared with global Chern or edge-state classifications in lattice models~\cite{LeeFuAng2024,LeeEtAl2025}.

\section{Berry curvature near the quadratic touching}\label{sec:berry_results}
The linear and quadratic terms of the effective Hamiltonians also predict distinct momentum profiles of the Berry curvature. A gapped linear Dirac cone concentrates curvature at the valley center. The detuned $AC$ Hamiltonian contains the factor $q^2$ in the numerator of Eq.~\eqref{eq:OmegaACeff}; its smooth Berry curvature vanishes at $q=0$ for nonzero $M_{AC}$ and reaches maximal magnitude at a finite radius.

\begin{figure*}[t]
\includegraphics[width=0.98\textwidth]{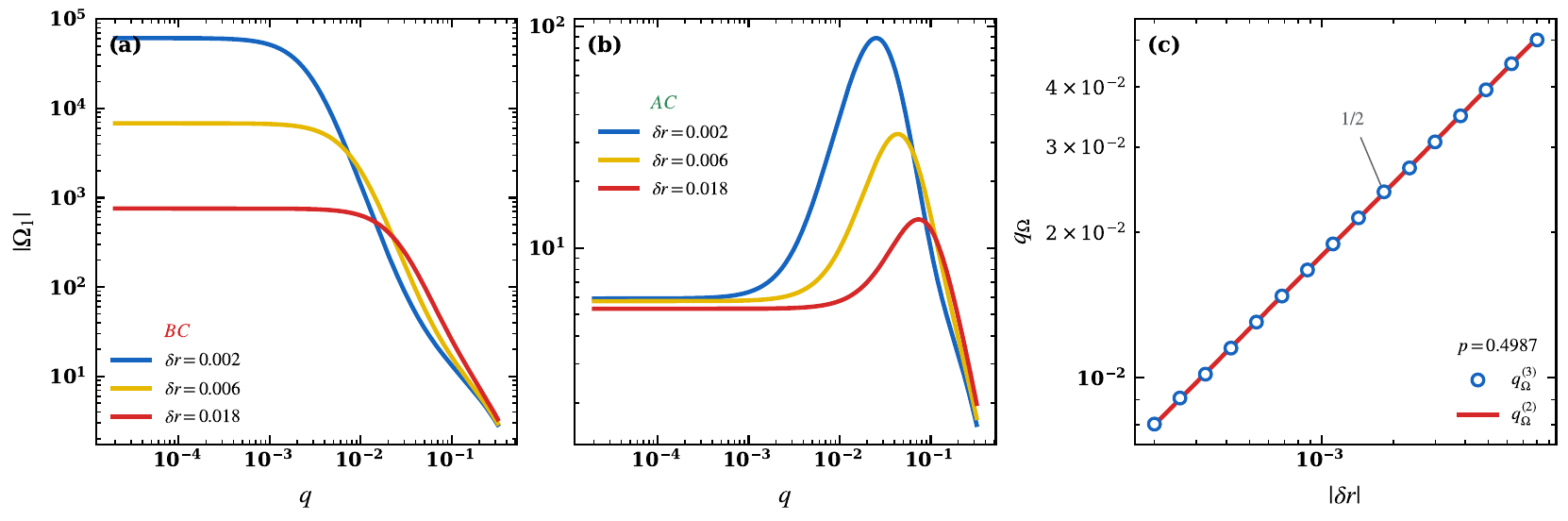}
\caption{Berry curvature near the linear $BC$ and quadratic $AC$ degeneracies at $\alpha=0.30$. (a) $|\Omega_1|$ for three $BC$ drive detunings. (b) Corresponding $AC$ profiles, with the maximum at finite $q$. (c) Curvature-ring radius from the complete three-band Kubo calculation and from Eq.~\eqref{eq:qOmega_exact}. The numerical points are fitted to a power law; the solid curve is Eq.~\eqref{eq:qOmega_exact} with coefficients fixed by the Schrieffer-Wolff reduction, with no fit to the three-band data.}
\label{fig:berry}
\end{figure*}

Figure~\ref{fig:berry} compares these predictions with the interband Kubo curvature of the complete three-band Hamiltonian. Panel (a) shows $|\Omega_1|$ for the linear $BC$ contact at three drive detunings: the profile remains centered at the valley and narrows as the mass is reduced. Panel (b) shows the corresponding $AC$ profiles, which vanish at $q=0$ and develop an annular maximum whose radius moves inward as the spectral detuning decreases. Panel (c) compares the numerically extracted radius with Eq.~\eqref{eq:qOmega_exact}. The extraction underlying the points in panel (c) uses only the full three-band data: at each drive detuning the interband Kubo formula is evaluated on a radial momentum grid of $500$ points spanning $q\in[10^{-4},2]$ (in units of $\lambda/\hbar v_F$), and after energy ordering the three bands provide three curvature magnitudes $|\Omega_n(q)|$ at every $q$. The band carrying the largest curvature magnitude is selected, and the ring radius is identified with the momentum at which $|\Omega_n|$ on that band attains its maximum. Every coefficient in the solid theoretical curve follows from the Schrieffer-Wolff reduction; no amplitude or exponent is adjusted to the three-band data. The full calculation approaches the square-root law $q_\Omega\propto|r-r_{AC}|^{1/2}$ in the low-energy interval. The finite-momentum maximum is therefore not a numerical accident: it arises from the competition between the quadratic mass detuning and the winding-two off-diagonal coupling proportional to $q^2$.

The exact valley relation in Eq.~\eqref{eq:valley_conjugation} provides a stringent symmetry constraint on the transport calculation. Opposite-valley spectra are reflected about zero energy, and the energy-ordered Berry curvatures reverse sign. The charge Hall conductivity is consequently even in $\mu$, while the charge transverse thermoelectric conductivity is odd, Eqs.~\eqref{eq:hall_even} and \eqref{eq:nernst_odd}. Figure~\ref{fig:Sparity} in Appendix~\ref{app:valleyparity} verifies both relations numerically over the transport window.

\section{Transverse thermoelectric inference}\label{sec:inverse_results}
The inverse analysis uses the charge coefficient $\alpha_{xy}^{c}$ summed over both valleys and both real-spin sectors. The Kubo response is evaluated on a grid in drive and chemical potential and then convolved with a Gaussian of width $\sigma_\mu$ to represent finite chemical-potential resolution. For scale extraction only, the reference response includes a $1\%$ additive perturbation; this term does not enter the Hamiltonian. The numerical conventions are summarized in Appendix~\ref{app:units}.

\begin{figure*}[t]
\includegraphics[width=0.98\textwidth]{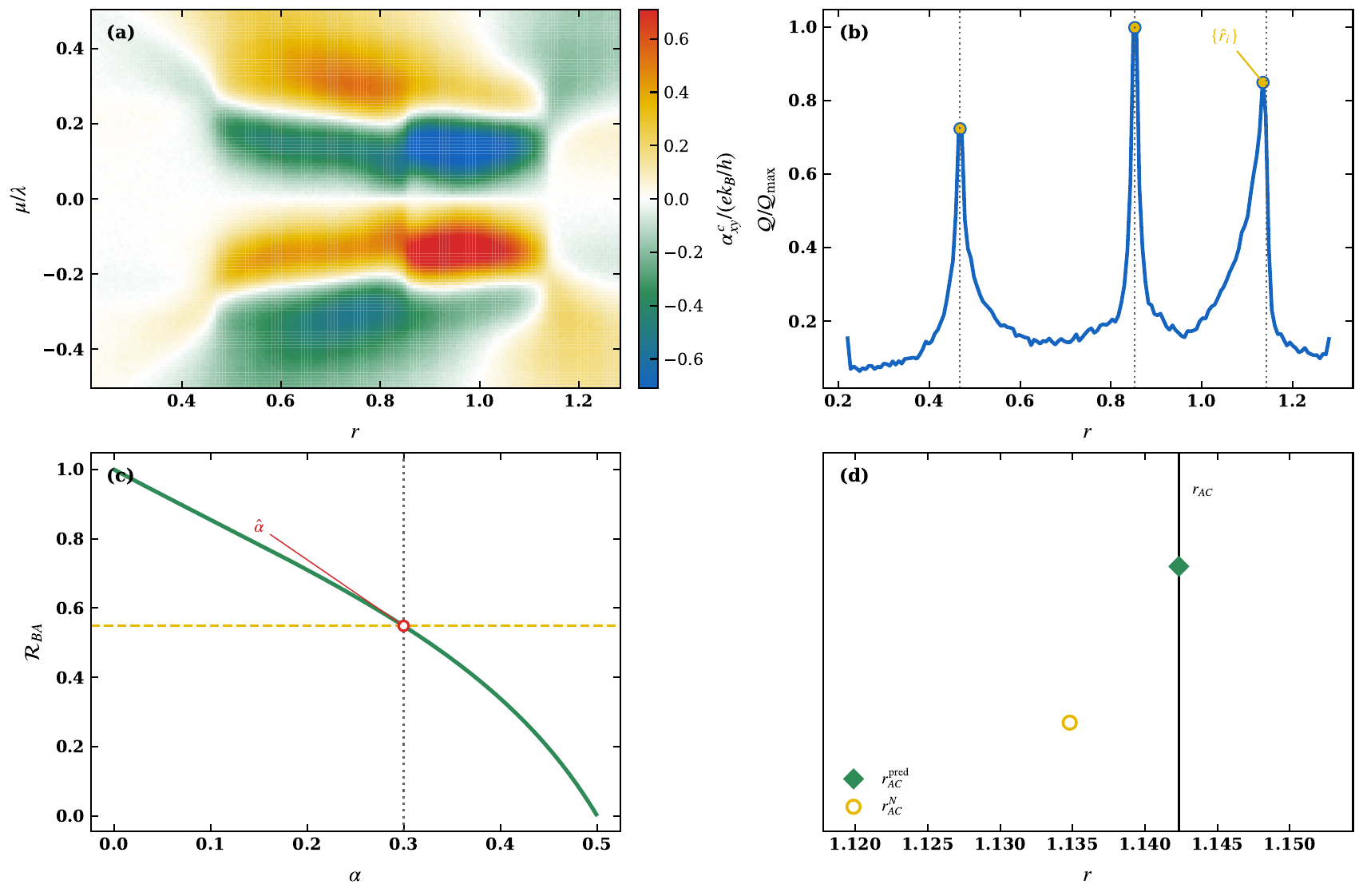}
\caption{Inference from the total charge transverse thermoelectric conductivity at $\alpha=0.30$. (a) Response including the finite-$\mu$ resolution kernel and the reference one-percent perturbation; the color scale is saturated only for display. (b) Normalized $\mathcal Q(r)$ and resolved response extrema; dotted lines show the exact spectral degeneracies. (c) Scale-free ratio $\mathcal R_{BA}(\alpha)$ and the value selected by the two linear response scales. (d) Exact $AC$ spectral degeneracy, prediction from the linear pair, and finite-temperature $AC$ response extremum.}
\label{fig:inverse}
\end{figure*}

Figure~\ref{fig:inverse}(a) shows the reference response map and panel (b) the corresponding change measure $\mathcal Q$ from Eq.~\eqref{eq:Q}. In the low-$\alpha$ interval, the first two resolved structures are associated with the linear $BC$ and $AB$ branches according to their analytic ordering. Their ratio intersects the monotone function $\mathcal R_{BA}(\alpha)$ in panel (c), yielding $\widehat\alpha$. The same two scales fix the overall drive calibration and therefore predict the $AC$ spectral degeneracy. The dotted spectral lines in panel (b) are shown for comparison.

Panel (d) separates the predicted spectral value from the $AC$ thermoelectric extremum extracted from the response. The spectral prediction is controlled by the two linear scales and the exact branch geometry. The thermoelectric extremum is controlled by an entropy-weighted momentum integral over the detuned quadratic spectrum and is systematically displaced. Treating those two quantities separately avoids conflating a spectral zero with the maximum of a finite-temperature response.

At the reference point, the scale ratio yields $\widehat\alpha\simeq0.2997$. One-percent response perturbations produce a standard deviation of about $1.5\times10^{-4}$, consistent with Eq.~\eqref{eq:errorprop}; the displacement of the finite-temperature $AC$ extremum is much larger. Appendix~\ref{app:uncertainty} summarizes this uncertainty estimate.

\section{Predictive domain}\label{sec:predictive}
\begin{figure*}[t]
\includegraphics[width=0.98\textwidth]{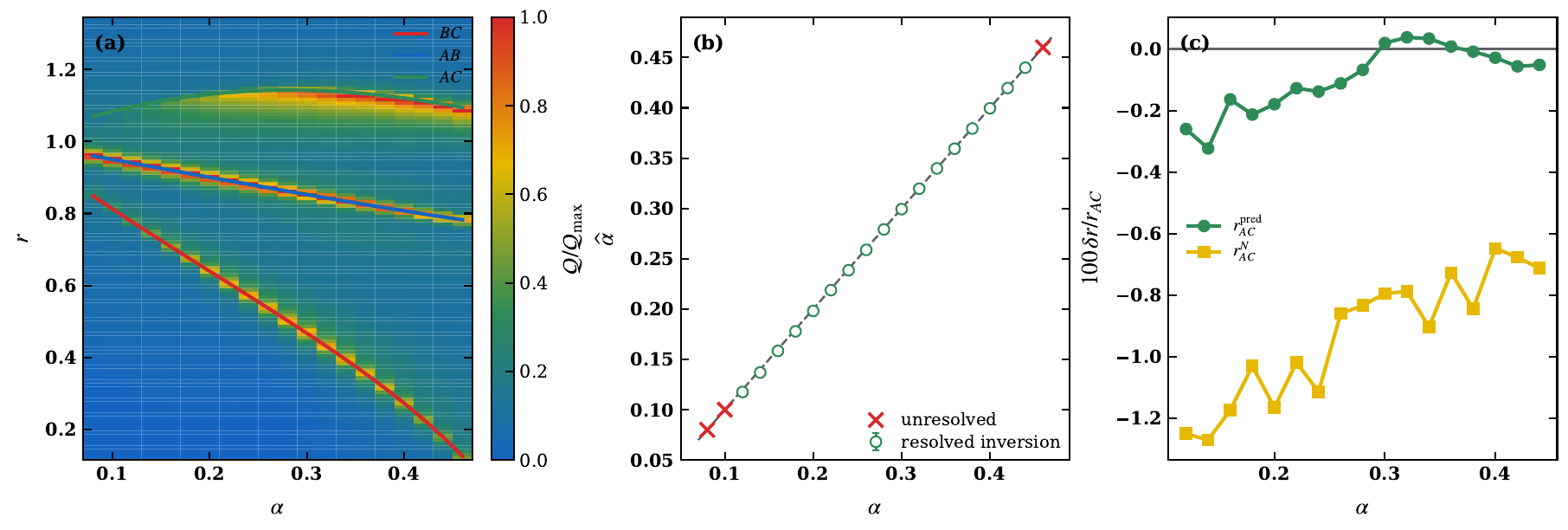}
\caption{Predictive domain of the two-linear-feature inversion. (a) Normalized response-change map $\mathcal Q(\alpha,r)$ with the analytic spectral degeneracies superposed. (b) Inferred $\widehat\alpha$ from forward response maps; error bars denote the response-perturbation standard deviation, and crosses mark unresolved cases. (c) Relative error of the predicted $AC$ spectral contact and displacement of the finite-temperature $AC$ thermoelectric extremum. The same $q$-grid, $\mu$-grid, Gaussian broadening, derivative procedure, peak-prominence criterion, and drive grid are used for all $\alpha$.}
\label{fig:predictive}
\end{figure*}

We repeat the response analysis over a dense set of interpolation parameters restricted to $\alpha<1/2$. Figure~\ref{fig:predictive}(a) shows $\mathcal Q(\alpha,r)$ from the calculated charge-response maps. The three response ridges follow the $BC$, $AB$, and $AC$ spectral curves over most of the interval. Near the lower and upper edges, finite drive resolution merges neighboring structures and the inversion ceases to be identifiable.

Panel (b) compares the inferred $\widehat\alpha$ with the generating value. The resolved points follow the diagonal without retuning the response analysis. Crosses mark cases in which the response structures merge or their ordering becomes ambiguous. Panel (c) compares the predicted $AC$ spectral contact with the finite-temperature $AC$ response position. The spectral prediction stays close to the Hamiltonian degeneracy throughout the identifiable interval; the response extremum exhibits a larger, parameter-dependent shift associated with the evolving quadratic dispersion and Berry-curvature profile.

The charge-response mapping changes after the $BC$ spin reassignment. Calculations above $\alpha=1/2$ exhibit additional or displaced extrema in the spin- and valley-summed response, so local thermoelectric maxima no longer map one-to-one onto the three pairwise $q=0$ degeneracies. Figure~\ref{fig:Sfail} in Appendix~\ref{app:noise} gives explicit examples. The quantitative inverse relation is consequently stated only over the interval supported by the forward calculations.

\section{Numerical control and physical range}\label{sec:scope}
The calculation is applicable to the regime in which an effective thermal distribution of the high-frequency effective Hamiltonian can approximate the electronic distribution; it is not a general nonequilibrium Floquet treatment. Within this regime, the numerical implementation satisfies three analytic checks: the Kubo curvature reduces to the graphene and dice endpoint expressions in Appendix~\ref{app:endpointberry}, the three band curvatures sum to zero away from degeneracies, and valley conjugation recovers the transport parities in Appendix~\ref{app:valleyparity}. The exact $AC$ factorization fixes the flat band and the quadratic dispersion directly, while the full-spectrum gap fits recover linear $AB/BC$ and quadratic $AC$ scaling. The thermoelectric calculation also approaches the Mott relation at low temperature (Appendix~\ref{app:mott}).

Appendix~\ref{app:convergence} summarizes the mesh and cutoff tests; among these numerical controls, the finite-drive grid most strongly affects the broader $AC$ response structure. The low-$\alpha$ branch assignment is therefore restricted to the sector in which the $BC$ branch keeps its spin sector; outside it the one-to-one correspondence with the three $q=0$ degeneracies is lost, as illustrated in Appendix~\ref{app:noise}.

The continuum calculation is controlled by the hierarchy in Eq.~\eqref{eq:epsA} and by momenta close to the valleys. It determines local band degeneracies, their low-energy dispersions, and the associated continuum Berry response. Integer Chern numbers and edge spectra require a compact Brillouin zone and are provided by lattice analyses of related $\alpha$-$T_3$ Hamiltonians~\cite{WangLiu2021,LeeFuAng2024,LeeEtAl2025}. The Fermi occupation used in Eqs.~\eqref{eq:sigma} and \eqref{eq:alphaN} corresponds to an effective-equilibrium off-resonant regime, such as a prethermal or bath-stabilized state~\cite{HoMoriAbaninDallaTorre2023}. A general driven steady-state distribution involves Floquet kinetic physics beyond this approximation. This hierarchy identifies the regime represented by the transport calculation; the exact spectral relations follow directly from the continuum Hamiltonian.

Connecting these predictions to experiment requires care with the measured observable. A typical open-circuit Nernst experiment does not directly test the transverse thermoelectric conductivity $\alpha_{xy}$ considered here. In an experiment, the transverse electric field $E_y$ is measured after a temperature gradient is introduced along the longitudinal direction while the transverse charge current is restricted to disappear, $j_y=0$. Thus, the open-circuit situation links the measured Nernst signal to the thermoelectric and electrical conductivity tensors,
\begin{equation}
\mathbf{j}
=
\hat{\sigma}\mathbf{E}
+
\hat{\alpha}(-\nabla T)
=
0,
\end{equation}
which gives
$
\mathbf{E}
=
\hat{\sigma}^{-1}\hat{\alpha}\nabla T$.
Therefore, separate measurements of the longitudinal and transverse electrical conductivities as well as the transverse thermoelectric response are typically needed in order to extract $\alpha_{xy}$ experimentally. As long as the driven electronic system stays sufficiently near the effective-equilibrium regime assumed in our calculation, the drive-dependent structures in $\alpha_{xy}$ can thus be accessed through Nernst measurements under circularly polarized irradiation in the present scheme. The Floquet parameter $r=\Delta_F/\lambda$ is determined by the laser intensity, polarization, and frequency, and the $\mu$-dependent response can be mapped experimentally through gate control of the chemical potential. The underlying Floquet-induced band degeneracies can then be identified experimentally by tracking the response features associated with the $AB$, $BC$, and $AC$ compensation conditions as functions of drive strength. The degree to which an effective Fermi--Dirac distribution can be established depends on Floquet heating, carrier relaxation, and the finite duration of the driven state, all of which must be included for a quantitative comparison with experiment.

\section{Conclusion}\label{sec:conclusion}
We have studied how intrinsic spin-orbit coupling and circular driving interact in the $\alpha$-$T_3$ lattice and obtained a set of tunable band degeneracies that result from the compensation of the respective drive- and spin-dependent mass components. The spectral implications of the three pairwise compensation criteria differ qualitatively. Unlike the $AB$ and $BC$ conditions, the $AC$ condition is met by an exact dark state of the complete three-band Hamiltonian, because the outer sites are not directly coupled. Instead of a general two-band quadratic crossover, the $AC$ condition yields an exact flat band in contact with a quadratically dispersive band. This distinction does not depend on a two-band approximation; rather, it is obtained directly from the complete three-band spectrum.

The quadratic contact carries its own geometric signature. Its low-energy description includes a quadratic off-diagonal coupling with winding number two away from the $AC$ compensation point. As a result, the Berry curvature shifts from the valley center to a finite-momentum ring, the radius of which depends on the spectral detuning according to the expected square root. This finite-momentum structure follows directly from the local quadratic band geometry, consistent with the agreement between the Berry curvature derived from the whole three-band Hamiltonian and the effective theory.

The calculations also indicate that the intrinsic transverse thermoelectric response retains signatures of these spectral structures. The two response features linked to the linear $AB$ and $BC$ connections define two different spectral scales in the interval $0<\alpha<1/2$. An unambiguous inverse mapping from the transverse thermoelectric response to the underlying band parameters is provided by their ratio, which varies monotonically with $\alpha$ and is independent of the overall spin-orbit energy scale. The location of the $AC$ flat- and quadratic-band degeneracy can be predicted without the need for its own response extremum if $\alpha$ and the common spectral scale are derived from these two features. Significantly, the thermoelectric extremum linked to the $AC$ feature typically differs from the real spectrum degeneracy due to the combined effects of momentum, Berry curvature, and entropy weighting in the transport integral.

The analytical degeneracy conditions, Berry-curvature structure, symmetry relations, and inverse mapping are reproduced by the complete three-band Kubo computations. The inversion has a finite domain of practical validity: the uniqueness of the inverse mapping is gradually reduced by response broadening, spin-sector reassignment of the $BC$ degeneracy, and finite chemical-potential resolution. Therefore, rather than directly identifying a thermoelectric extremum with a band crossing, the suggested method should be seen as an inverse characterization of the intrinsic response of the effective driven system.

In general, our findings show how the tunable geometry of a multiband spectrum can be transformed into experimentally accessible thermoelectric signatures through Floquet--spin--orbit engineering. The flat- and quadratic-band contact's precise dark-state mechanism results from the $\alpha$-$T_3$ model's three-site connectivity and the lack of direct coupling between the outer sites, rather than being an artifact of the specific parameter values. This implies that other multiband systems with comparable connectivity may host similar compensation-induced flat-band and nonstandard band-touching structures. Beyond the effective-equilibrium high-frequency description employed here, lattice corrections, disorder and scattering, Floquet heating, and nonequilibrium occupations will also need to be taken into account for a quantitative experimental implementation. Together, these results link inverse thermoelectric characterization to multiband band degeneracies, Berry-curvature geometry, and Floquet engineering in one driven platform.

\section*{Acknowledgments}
IK and GX acknowledge financial support from the NSFC under Grant No. 12674335.

\appendix
\makeatletter
\@addtoreset{figure}{section}
\@addtoreset{equation}{section}
\makeatother
\renewcommand{\thefigure}{\Alph{section}\arabic{figure}}
\renewcommand{\theequation}{\Alph{section}\arabic{equation}}
\setcounter{figure}{0}
\setcounter{equation}{0}

\section{Hamiltonian convention and numerical units}\label{app:units}
This appendix records the numerical conventions used in the figures below. The calculations diagonalize the continuum Hamiltonian of Sec.~\ref{sec:model} with valley $\eta$, real spin $s$, and optical helicity $\xi$ kept as independent labels. Energies are measured in units of the continuum spin--orbit scale $\lambda$, and momenta in units of $\lambda/(\hbar v_F)$. The positive optical coordinate is $r=\Delta_F/\lambda$, with
\begin{equation}
 \Delta_F=\frac{(e v_F A_0)^2}{(1+\alpha^2)\hbar\omega}.
\end{equation}
Thus the spectra and transport functions shown below are dimensionless functions of $r$; conversion to laser intensity is material and frequency-dependent.

Unless stated otherwise, the reference transport calculation uses $\alpha=0.30$, $k_BT/\lambda=0.035$, and a Gaussian chemical-potential resolution $\sigma_\mu/\lambda=0.020$. All momentum cutoffs and quadrature meshes are chosen so that the reported response scales are converged on the drive grid; explicit meshes and numerical settings are provided with the executable source. A small additive perturbation is used only when quoting response-extraction uncertainties and does not enter the Hamiltonian or the Berry curvature.

\section{Exact $AC$ structure and local phase space}\label{app:acfactor}
At the $AC$ degeneracy, $m_A=m_C\equiv E_c$. The characteristic polynomial of the complete three-band Hamiltonian then factorizes,
\begin{equation}
 \det(E-H)=(E-E_c)\left[(E-E_c)(E-m_B)-|\gamma_\eta|^2\right].
 \label{eq:Sfactor}
\end{equation}
One eigenvalue is therefore pinned to $E_c$ for all continuum momenta. The other two are
\begin{equation}
 E_{\pm}=E_c+\frac{-\Delta_B\pm\sqrt{\Delta_B^2+4|\gamma_\eta|^2}}{2},
 \qquad \Delta_B=E_c-m_B .
\end{equation}
The corresponding flat-band eigenstate may be written
\begin{equation}
 |D_\eta\rangle=
 \begin{pmatrix}
 u\gamma_\eta/|\gamma_\eta|\\[2pt]0\\[2pt]-c\gamma_\eta^*/|\gamma_\eta|
 \end{pmatrix},
 \qquad H|D_\eta\rangle=E_c|D_\eta\rangle .
\end{equation}
Its vanishing $B$-sublattice weight is the destructive-interference condition for the $A$--$B$--$C$ chain. Integrating the Berry connection around a fixed-$q$ loop encircling one valley gives
\begin{equation}
 \gamma_D=2\pi\eta\frac{1-\alpha^2}{1+\alpha^2}\pmod{2\pi}.
\end{equation}
The branch that meets the flat eigenvalue has the expansion
\begin{equation}
 E_q-E_c=\frac{|\gamma_\eta|^2}{\Delta_B}
 -\frac{|\gamma_\eta|^4}{\Delta_B^3}+O(q^6),
\end{equation}
which fixes both the quadratic leading dispersion and its $q^4$ correction without a fitting ansatz.

\begin{figure*}[ht!]
\includegraphics[width=0.94\textwidth]{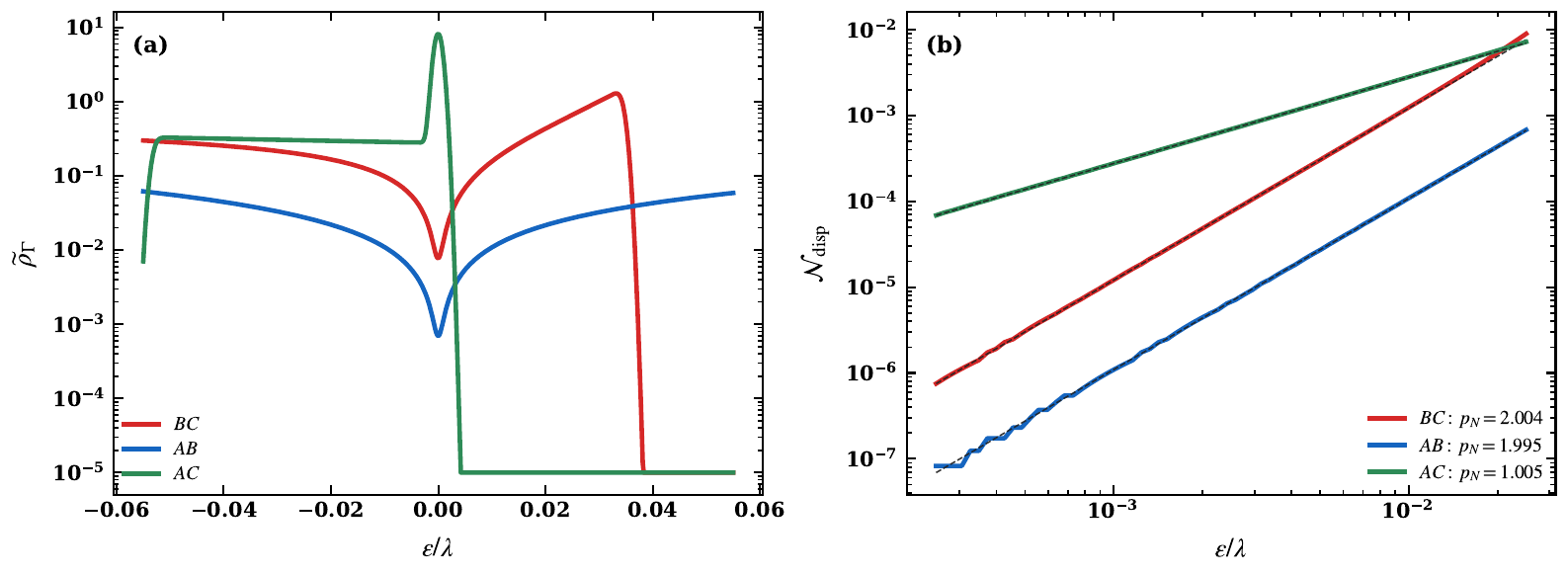}
\caption{Local spectral phase space at the three critical drives for $\alpha=0.30$. (a) Broadened spectral weight $\widetilde\rho_{\Gamma}(\varepsilon)$ from the three energy-ordered eigenvalues in a common momentum patch. The narrow $AC$ contribution contains the exact flat band. (b) Cumulative dispersive phase space $\mathcal N_{\rm disp}$. The low-energy powers approach $2$ for the linear $BC/AB$ contacts and $1$ for the dispersive $AC$ branch.}
\label{fig:Sphase}
\end{figure*}

The low-energy density of states consequently contains a flat-band singularity and the one-sided constant density of states of the quadratic branch,
\begin{equation}
 \rho_{AC}(E)=W_{\rm flat}\delta(E-E_c)
 +C_{AC}\Theta[(E-E_c)\Delta_B]+\cdots .
\end{equation}
For comparison with the full eigenspectrum, define the broadened local spectral weight in a radial patch $0\le q\le q_c$ by
\begin{equation}
 \widetilde\rho_{\Gamma}(\varepsilon)=
 \sum_{n=1}^{3}\int_0^{q_c}q\,dq\,
 \frac{\exp\{-[\varepsilon-(E_n(q)-E_c)]^2/(2\Gamma_E^2)\}}
 {\sqrt{2\pi}\Gamma_E},
 \label{eq:Slocdos}
\end{equation}
and the cumulative dispersive phase space by
\begin{equation}
 \mathcal N_{\rm disp}(\varepsilon)=
 \sum_{n\in\mathcal D}\int_0^{q_c}q\,dq\,
 \Theta\!\left(\varepsilon-|E_n(q)-E_c|\right).
 \label{eq:Scumulative}
\end{equation}
At $AC$ the exactly flat branch is retained in $\widetilde\rho_\Gamma$ but excluded from $\mathcal N_{\rm disp}$. Figure~\ref{fig:Sphase} shows the resulting phase space. The cumulative slopes distinguish the two linear contacts (power 2) from the quadratic dispersive branch (power 1) without using the effective-theory exponents as fitting inputs.

\section{Endpoint Berry curvature}\label{app:endpointberry}
\begin{figure*}[tbp]
\includegraphics[width=0.88\textwidth]{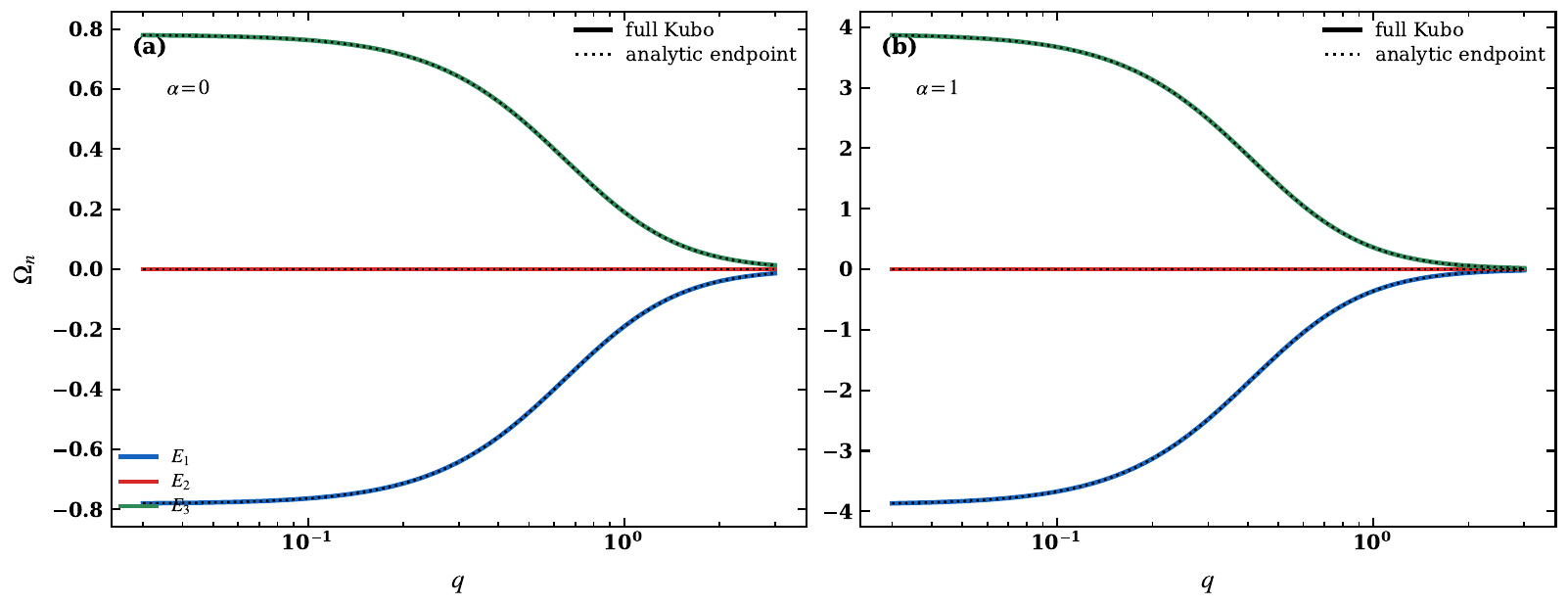}
\caption{Endpoint Berry curvature at (a) $\alpha=0$ and (b) $\alpha=1$. Blue, red, and green denote the energy-ordered bands $E_1$, $E_2$, and $E_3$, respectively. Solid curves are the complete three-band Kubo result; dotted curves are the analytic endpoint expressions of Eq.~\eqref{eq:endpointberry}.}
\label{fig:Sberry}
\end{figure*}

At the graphene and dice endpoints, the Kubo curvature reduces to the massive pseudospin-$1/2$ and pseudospin-1 forms. For the energy-ordered lower and upper bands,
\begin{equation}
 \Omega_{1,3}(q)=\mp\eta\,\nu_\alpha
 \frac{M v_F^2}{\left(M^2+v_F^2q^2\right)^{3/2}},
 \label{eq:endpointberry}
\end{equation}
with $\nu_\alpha=1/2$ at the graphene endpoint $\alpha=0$ and $\nu_\alpha=1$ at the dice endpoint $\alpha=1$; the middle-band curvature vanishes. Figure~\ref{fig:Sberry} compares Eq.~\eqref{eq:endpointberry} with the interband Kubo curvature of the complete $3\times3$ Hamiltonian.

\section{Valley conjugation and transport parity}\label{app:valleyparity}
\begin{figure*}[ht!]
\includegraphics[width=0.88\textwidth]{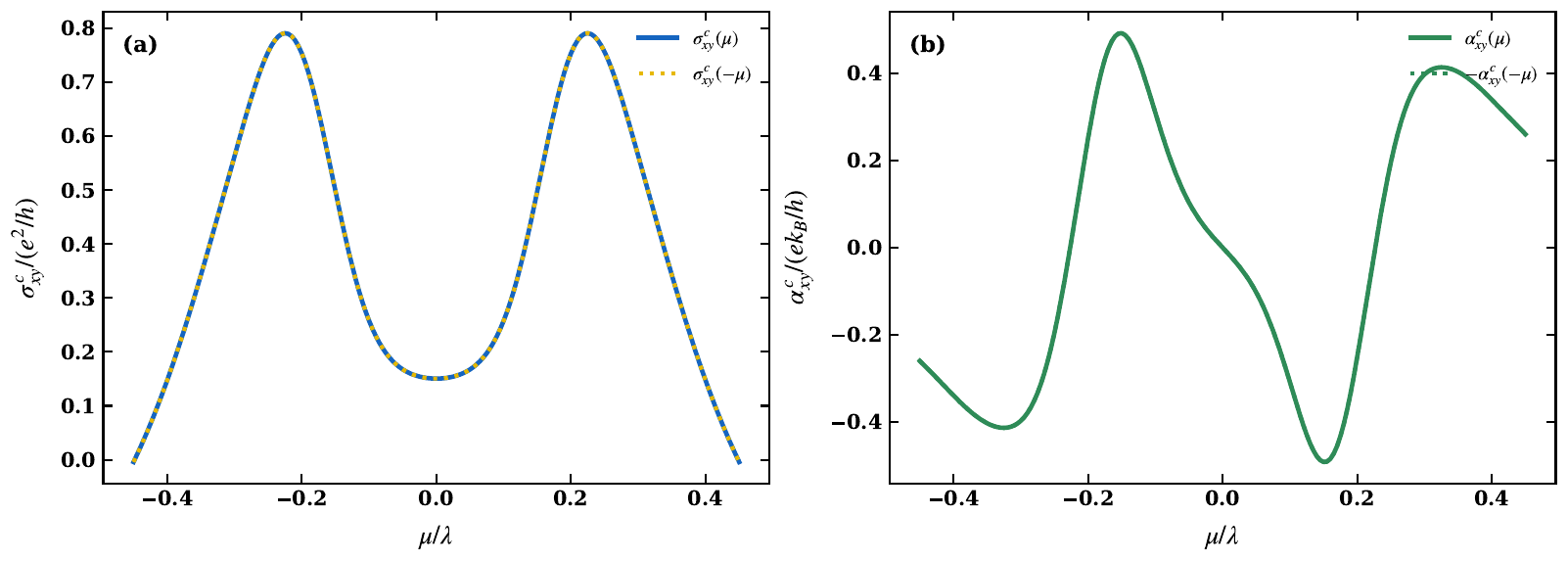}
\caption{Valley-summed transport parity. (a) Blue solid curve, $\sigma_{xy}^{c}(\mu)$; yellow dotted curve, $\sigma_{xy}^{c}(-\mu)$. (b) Green solid curve, $\alpha_{xy}^{c}(\mu)$; light-green dotted curve, $-\alpha_{xy}^{c}(-\mu)$. The exact overlap of each solid curve with its dotted symmetry transform confirms valley conjugation.}
\label{fig:Sparity}
\end{figure*}

The exact valley relation $H_{-\eta}=-H_\eta^*$ implies
\begin{align}
 E_{n,-\eta}(\mathbf q)&=-E_{4-n,\eta}(\mathbf q),\\
 \Omega_{n,-\eta}(\mathbf q)&=-\Omega_{4-n,\eta}(\mathbf q).
\end{align}
After summing valleys and spins, these identities give an even charge Hall conductivity and an odd charge transverse thermoelectric conductivity,
\begin{equation}
 \sigma_{xy}^{c}(\mu)=\sigma_{xy}^{c}(-\mu),\qquad
 \alpha_{xy}^{c}(\mu)=-\alpha_{xy}^{c}(-\mu).
\end{equation}
Figure~\ref{fig:Sparity} shows the directly calculated responses and their symmetry transforms.

\section{Low-temperature Mott limit}\label{app:mott}
The entropy-weighted thermoelectric integral must reduce to the low-temperature Mott form of the Hall response. This relation provides an independent check on the low-temperature limit of the numerical thermoelectric integration; it is used as a validation test rather than displayed as a separate figure. In the dimensionless convention used in the numerical work, Eq.~\eqref{eq:Mott} is evaluated as
\begin{equation}
 \frac{\alpha_{xy}^{c}}{e k_B/h}
 \simeq-\frac{\pi^2}{3}\frac{k_BT}{\lambda}
 \frac{\partial}{\partial(\mu/\lambda)}
 \left(\frac{\sigma_{xy}^{c}}{e^2/h}\right).
\end{equation}
The direct entropy integral approaches this derivative form as $T$ is lowered; the numerical values of this comparison are retained in the reproducibility package.

\section{Uncertainty of the two-scale inversion}\label{app:uncertainty}
The parameter uncertainty is controlled by the two measured linear scales rather than by the broader $AC$ response feature. Defining $R=D_{BC}/D_{AB}$ and letting $C$ be the covariance matrix of $\mathbf D=(D_{BC},D_{AB})$, linear propagation gives Eq.~\eqref{eq:errorprop}. At the reference point $\alpha=0.30$, the response calculation yields $\widehat\alpha\simeq0.2997$, and a one-percent response perturbation gives $\sigma_\alpha\simeq1.5\times10^{-4}$. This uncertainty is well below the finite-temperature displacement of the $AC$ thermoelectric extremum from the predicted spectral contact.

\section{Numerical implementation and convergence}\label{app:convergence}
The complete $3\times3$ Hamiltonian is diagonalized on a radial momentum mesh. The Berry curvature is computed from the interband Kubo formula, and the Hall and thermoelectric coefficients are obtained by direct radial integration. The production meshes were checked against finer momentum resolution, larger momentum cutoff, denser chemical-potential grid, and finer drive grid. The first three controls are already converged on the production meshes; the finite drive spacing is the dominant remaining uncertainty for the broader $AC$ response. Numerical settings, convergence tables, source code, and a self-contained Jupyter notebook are included in the reproducibility package.

\section{Breakdown of the low-$\alpha$ response assignment}\label{app:noise}
\begin{figure*}[ht!]
\includegraphics[width=0.92\textwidth]{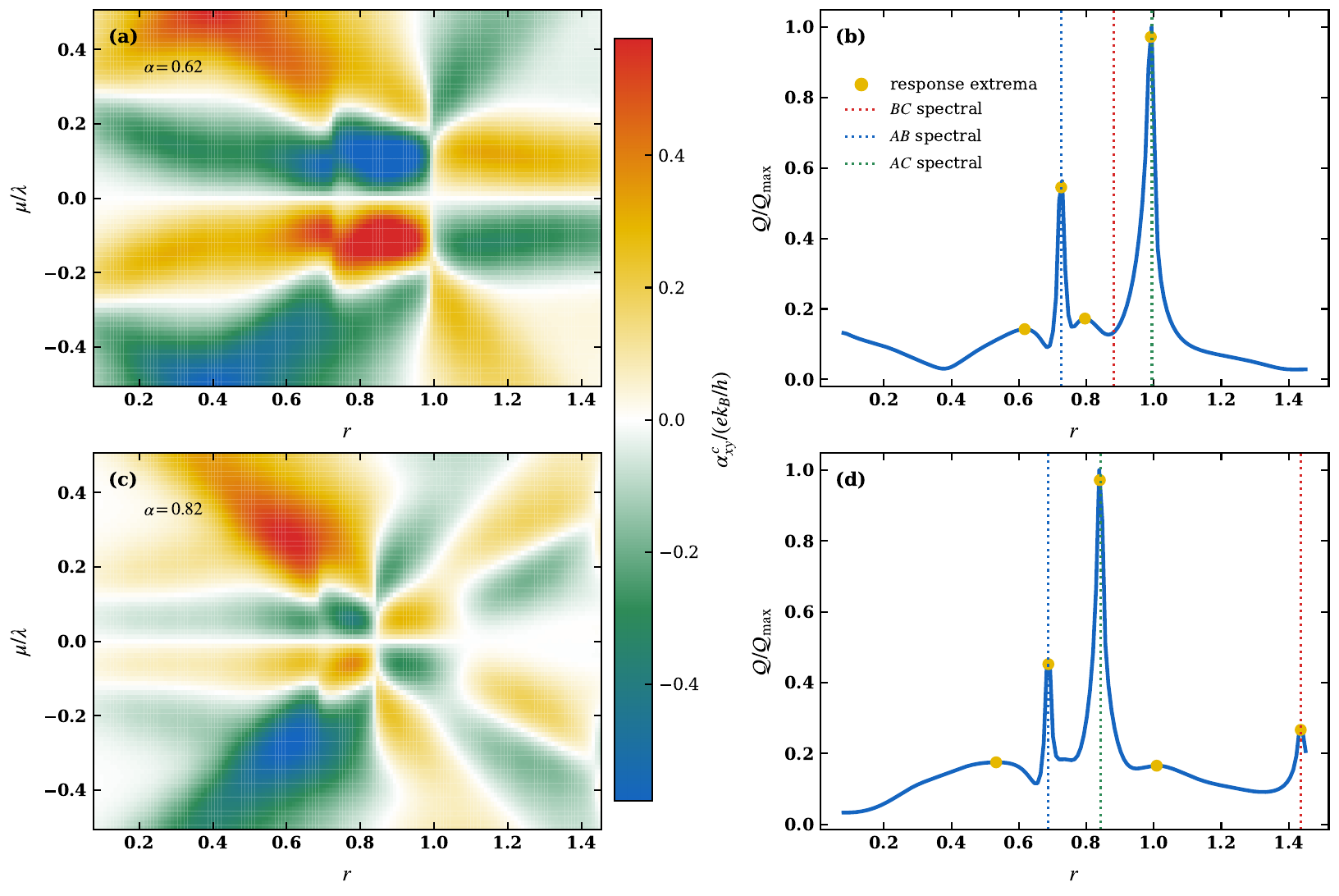}
\caption{Breakdown of the low-$\alpha$ inversion outside its sector. (a,c) Forward charge transverse thermoelectric maps at $\alpha=0.62$ and $0.82$; the shared color bar gives $\alpha_{xy}^{c}/(e k_B/h)$. (b,d) Normalized response-change measure $\mathcal Q/\mathcal Q_{\max}$. Yellow circles mark response extrema. Dotted vertical lines mark the $q=0$ spectral degeneracies obtained from the Hamiltonian: red for $BC$, blue for $AB$, and green for $AC$. The mismatch between response extrema and spectral lines shows why the low-$\alpha$ inversion is not extrapolated through the spin-reassignment regime.}
\label{fig:Sfail}
\end{figure*}

The two-linear-scale inversion is tied to the branch ordering of the $0<\alpha<1/2$ sector. After the $BC$ spin reassignment, the spin- and valley-summed thermoelectric response develops additional or displaced extrema even though the Hamiltonian degeneracies remain well defined. Figure~\ref{fig:Sfail} gives two representative forward calculations. The figure therefore marks the physical boundary of the inverse construction.

\begin{widetext}
\section{Full-band energies and bandwise dispersive shifts}\label{app:bandshifts}
\begin{figure*}[ht!]
\includegraphics[width=0.94\textwidth]{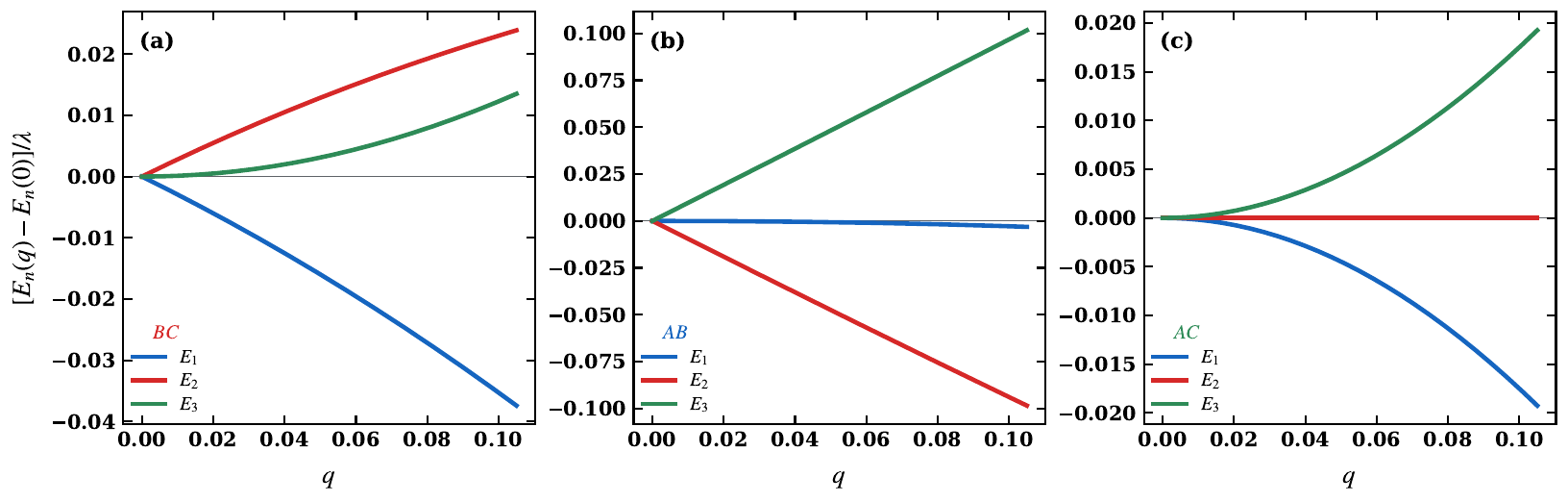}
\caption{Bandwise dispersive shifts at the three critical drives for $\alpha=0.30$. Each panel shows $[E_n(q)-E_n(0)]/\lambda$ obtained by direct diagonalization of the complete Hamiltonian, with blue, red, and green denoting $E_1$, $E_2$, and $E_3$. Unlike Fig.~\ref{fig:dispersion}(d--f), which uses a common crossing energy $E_c$ for each panel, this representation isolates the momentum dependence of each energy-ordered band.}
\label{fig:Sbandshift}
\end{figure*}
The lower panels of Fig.~\ref{fig:dispersion} use a single crossing energy $E_c$ for each pair and plot $(E_n-E_c)/\lambda$. That representation keeps the nonparticipating third band visible. A different quantity isolates the momentum-induced dispersion of each band,
\begin{equation}
 \delta E_n(q)=E_n(q)-E_n(0).
 \label{eq:Sbandshift}
\end{equation}
Because Eq.~\eqref{eq:Sbandshift} subtracts a different reference energy from each band, all three curves start at zero even when one band is remote from the crossing energy. Figure~\ref{fig:Sbandshift} shows this complementary representation from the same complete $3\times3$ diagonalization. It is useful for displaying the small remote-band dispersion, but it is not the energy convention used to define the critical gap in Fig.~\ref{fig:dispersion}.
\end{widetext}
\section*{Data and code availability}
Available upon request from the corresponding author.

\makeatletter
\if@filesw
 \immediate\write\@auxout{\string\citation{REVTEX42Control}}%
\fi
\makeatother

\bibliographystyle{apsrev4-2}
\bibliography{refs}
\end{document}